\documentclass[twocolumn,
nofootinbib,superscriptaddress,aps,10pt,longbibliography]{revtex4-1}

\usepackage{bbold}

\usepackage[usenames,dvipsnames]{xcolor}
\usepackage{amsmath}
\usepackage{amsthm, amssymb}
\usepackage{enumitem}
\usepackage{slashed}
\usepackage{graphicx}
\usepackage{tikz}
\usetikzlibrary{calc,fadings,decorations.pathreplacing,shapes,shapes.multipart,arrows,shapes.misc,intersections,positioning,patterns}
\usepackage{bm}
\usepackage{dsfont}
\usepackage{changepage}
\usepackage{array}
\usepackage{appendix}
\usepackage{subfigure}

\definecolor{bluepurple2}{rgb}{0.06,0,0.6}
\usepackage[colorlinks=true,citecolor=blue,linkcolor=bluepurple2]{hyperref}

\newcommand{\seq}{s_{\mathrm{eq}}}

\newcommand{\up}{\mathord{\uparrow}}

\usepackage{bm}

\graphicspath{{./}{./images/}}

\newcommand{\bit}{\begin{itemize}}
\newcommand{\eit}{\end{itemize}}

\newcommand{\f}{\frac}
\renewcommand{\>}{\right\rangle}
\newcommand{\<}{\left\langle}
\newcommand{\ba}{\begin{align}}
\newcommand{\ea}{\end{align}}
\newcommand{\be}{\begin{equation}}
\newcommand{\ee}{\end{equation}}
\newcommand{\bi}{\begin{itemize}}
\newcommand{\ei}{\end{itemize}}
\newcommand{\lf}{\left(}
\newcommand{\ri}{\right)}
\newcommand{\dd}{\mathrm{d}}

\newcommand{\Tr}{\operatorname{Tr}}

\newcommand{\red}{\color{red}}
\newcommand{\blue}{\color{blue}}

\begin{document}
\date{\today}

\newcommand{\bra}[1]{\< #1 \right|}
\newcommand{\ket}[1]{\left| #1 \>}

\newcommand{\comment}[1]{{\blue [#1]}}
\newcommand{\rem}[1]{{\red [#1]}}
\newcommand{\rough}[1]{{\textcolor{gray}{#1}}}

\newcommand{\ra}{\rangle}
\newcommand{\la}{\langle}

% \title{Continuum model for entanglement growth in noisy interacting fermion chains}
\title{Continuum mechanics of entanglement in noisy interacting fermion chains}

\author{Tobias Swann}
\affiliation{Rudolf Peierls Centre for Theoretical Physics, Clarendon Laboratory, Parks Road,
Oxford OX1 3PU, UK}

\author{Adam Nahum}
\affiliation{Laboratoire de Physique de l’\'Ecole Normale Sup\'erieure, CNRS, ENS \& Universit\'e PSL, Sorbonne Universit\'e, Universit\'e Paris Cit\'e, 75005 Paris, France}

\date{\today}

\begin{abstract}
We develop an effective continuum description for information scrambling in a chain of randomly-interacting Majorana fermions. The approach is based on the semiclassical treatment of the path integral for an effective spin chain that describes ``two-replica'' observables such as the entanglement purity and the OTOC. This formalism gives exact results for the entanglement membrane and for operator spreading in the limit of weak interactions. In this limit there is a large crossover lengthscale between free and interacting behavior, and this large lengthscale allows for a continuum limit and a controlled saddle-point calculation. The formalism is also somewhat different from that known from random unitary circuits. The entanglement membrane emerges as a kind of bound state of two travelling waves, and shows  an interesting unbinding phenomenon as the velocity of the entanglement membrane approaches the butterfly velocity.
\end{abstract}

\maketitle

\section{Introduction}

Solvable models for studying 
chaotic many-body dynamics are rare.
Chaotic field theories with a holographic dual provide one solvable regime \cite{blake2022snowmass}.
Minimal models based  on unitary circuits have proven useful for studying entanglement in generic many body systems and have led to putatively universal predictions \cite{fisher2023random,bertini2025exactly}. 
By taking noise averages over several replicas of a random unitary circuit, it is possible to express quantities such as the R\'enyi entanglement entropies as corresponding quantities in a classical statistical mechanics problem. However, random unitary circuits are generally strongly interacting, and the corresponding statistical mechanics problems are discrete problems on the lattice.

Here we consider a  different solvable regime, studying a model which is nearly free-fermion but has weak interactions.
The consequence of weak interactions is 
that there is a crossover length scale,
beyond which the characteristic interacting behavior emerges,
that is much larger than the lattice spacing. 
As a result we obtain an an effective \textit{continuum} model for entanglement, 
which can be treated in a controlled way using a semiclassical approximation~\cite{swann2023spacetimepictureentanglementgeneration}.

The weakly interacting fermion model is illuminating for several reasons. 
First, it confirms the universality of the ``entanglement membrane'' picture
\cite{jonay2018coarse,nahum2017quantum,mezei2018membrane,mezei2020exploring,Nahum_2018,Chan_2018,ZhouNahum,zhou2020entanglement,rakovszky2019entanglement,kudler2020quantum,agon2021bit,kudler2021entanglement,rampp2024entanglement,vardhan2026entanglement,brauner2022snowmass,jiang2025entanglement,bao2018entropy,mezei2020black}, for  dynamics of quantum information, in a new regime.
Second, the present continuum limit gives an alternative source of analytical control.
Third, and most interesting, it reveals structures that were not apparent in previous lattice treatments, and which we believe will be relevant to a wider class of models.

For simplicity we focus on an interacting fermion chain with no continuous symmetries (i.e. a Majorana chain) \cite{swann2023spacetimepictureentanglementgeneration}, and analyze quantities expressible with ${n=2}$ replicas of the circuit and its conjugate: the averages of entanglement purity and of an out-of-time-ordered correlator.  
The fermion model we study is  similar to an SYK chain \cite{chowdhury2022sachdev} with noisy couplings \cite{liu2021non,
zhang2021emergent,
agarwal2022emergent,
agarwal2023charge}, but our exact results do not require any ``large $N$'' limit: there is a single Majorana mode on each site. 
The model can also be viewed as an interacting, Majorana version of the quantum symmetric simple exclusion process \cite{Bauer_2017,Bauer_2019,bernard2022dynamics}.

The replicated fermion model is mapped to a Heisenberg spin chain that can be treated semiclassically,
following \cite{swann2023spacetimepictureentanglementgeneration}.
The semiclassical treatment of the path integral for the effective spin chain  involves two spacetime fields,  $z(x,t)$ and $\bar{z}(x,t)$.
By solving their equations of motion, we get a continuum description of the entanglement membrane as a bound state of two smooth domain walls, one in $z$  and one in $\bar{z}$.
We obtain a prediction for the entanglement membrane tension  $\mathcal{E}(v)$, which is a function of the velocity $v$ that sets the orientation of the membrane in spacetime,
in terms of solutions of differential equations.
We discussed the special case of a static membrane, $v=0$, previously \cite{swann2023spacetimepictureentanglementgeneration}: this case simplifies considerably, because the symmetries of the problem mean that  $\bar{z}(x,t)=z(x,t)^*$  and that the solution is time-independent, so that  the problem reduces to a simple classical energy minimisation problem for  $z(x)$. More generally, it is necessary to address the dynamical problem.
Previously the derivation of the entanglement membrane picture has been under quantitative control only in  holographic systems (where very complete results, including coupling to hydrodynamics, are available \cite{mezei2018membrane,mezei2020exploring}) 
and in quantum circuits (see e.g.~\cite{ZhouNahum,zhou2020entanglement,rampp2024entanglement,vardhan2026entanglement}).

In addition to a prediction for $\mathcal{E}(v)$, we also find a qualitative change at a critical velocity $v_c$ where the domain walls in $z$ and $\bar{z}$ become unbound.
As ${v\to v_c}$ from below, the characteristic size of the bound state diverges like ${1/\sqrt{v_c-v}}$.
In the unbound regime we find two domain walls that travel separately as travelling waves with fixed velocity $v_c$. 
This qualitative change at a critical velocity points to ballistic spreading of operators in the Majorana chain, and we identify the critical velocity $v_c$ with the butterfly velocity $v_B$.

It is interesting to compare the picture we get with the corresponding picture for the Haar-random circuit. There are many similarities, with an emergent entanglement membrane whose effective line tension $\mathcal{E}(v)$ depends on velocity (so that the basic universal features of the dynamics of entanglement and OTOCs agree at the largest lengthscales). However, there are also differences. The effective statistical mechanics model for the circuit can also be formulated in terms of two degrees of freedom which (like $z$ and $\bar{z}$) are related by time reversal \cite{Nahum_2018}, but in the circuit these are  summed over separately, 
whereas $z$ and $\bar{z}$ are not independent in the  path integral (even if we allow them to be independent when we search for a saddle point). 
In the circuit it is convenient to integrate out one of the two degrees of freedom, which simplifies the model at the cost of making time-reversal symmetry no longer manifest.
In our continuous model there is not an obvious analog and we treat $z$ and $\bar z$ symmetrically.

The present system illustrates how the entanglement structures of the noisy free fermion system (in which, for example, entanglement spreads diffusively) 
cross over to those of the interacting system.
This extends the analysis of our previous paper \cite{swann2023spacetimepictureentanglementgeneration}, which considered the crossover due to interactions  for the state entanglement after a quench from a product state.
A crossover induced by interactions has recently been analyzed \cite{Poboiko_2025,Guo_2025} in the case of monitored free fermions, using the nonlinear sigma model description \cite{jian2023measurement,sigmamodelmeasurement,poboiko2023theory,fava2024monitored}.
However the continuum theory is different in the unitary case and in the monitored case, and the techniques used here will be different. 

Travelling wave equations \cite{kolmogorov1study,fisher1937wave} play a role in one part of our the analysis.
Various kinds of  travelling wave equation have arisen in descriptions of  OTOCs  \cite{aleiner2016microscopic,xu2019locality,aron2023traveling,aron2023kinetics, zhou2025operator,chen2019quantum,zhou2023hydrodynamic,Nahum_2018,xu2019locality,keselman2020scrambling,delucaANunpublished}, including in a noisy fermion chain similar to ours but with U(1) charge conservation symmetry \cite{agarwal2023charge}.
In fact the recent Ref.~\cite{parrikar2025entanglement} studies a Majorana chain similar to ours, though with many flavours: while the observables and formalism differ from those here,  Ref.~\cite{parrikar2025entanglement}  arrives at a travelling wave equation equivalent to one that we encounter.

However, the description of the entanglement membrane  here involves new ingredients.
While the membrane  emerges as a 
solution of the saddle-point equations that propagates at a fixed velocity,\footnote{To avoid confusion, we note that we do not use the term ``travelling wave'' to refer to the bound-state solution, but only to  solutions in which only one of $z$ or $\bar z$ has nontrivial variation, and for which the equations of motion simplify to FKPP-like equations.} 
this solution is not like a conventional travelling wave, and the saddle-point equations do not in general reduce to conventional travelling wave equations.
Instead the entanglement membrane is a ``bound state'' of waves in two fields, $z$ and $\bar z$,  which are related by time-reversal symmetry and for which the natural ``arrows of time'' are opposite.

\section{Model}
\label{sec:model}

The model we study is a 1D fermionic Hamiltonian with 
quadratic (free-fermion) 
nearest-neighbour couplings 
as well as quartic interactions. 
For simplicity, 
both of these are taken to be noisy i.e. varying randomly in both space and time. 
By a Jordan-Wigner transformation, the model can also be mapped to a generalized Ising model with nearest and next-nearest-neighbour interactions.\footnote{With noisy terms of the forms $X_r$ and $Z_{r} Z_{r+1}$
(arising from the free-fermion couplings) 
and $X_{r}X_{r+1}$ and $Z_{r}Z_{r+2}$ (from the fermion interactions), where $r$ is the site index in the Ising chain.}

We also assume that the quartic interactions are very weak compared to the free-fermion couplings, as this allows us to derive a continuum model for the spreading of entanglement throughout the chain.
If the strength of the quartic interactions were set to exactly zero, then the system would be a free-fermion system, leading to qualitatively different behaviour of the entanglement growth that was discussed in Ref.~\cite{swann2023spacetimepictureentanglementgeneration}. 
However, for any non-zero strength of quartic interactions the system crosses over into the interacting regime at sufficiently long time scales.

We will define our system using a Majorana representation. We take a 1D chain of Majorana modes of length $L$ (with $L$ even). At each site $i=1,\dots,L$ we have a Majorana operator $\gamma_i$, where $\{\gamma_i,\gamma_j\}=2\delta_{ij}$. The time-dependent Hamiltonian is
\begin{equation} \label{eq:majoranahamiltonian}
    H_\gamma(t) = -i\sum_i \eta_i(t)\gamma_i\gamma_{i+1}
    - \sum_i \eta'_i(t)\gamma_i\gamma_{i+1}\gamma_{i+2}\gamma_{i+3},
\end{equation}
where $\eta_i(t)$ and $\eta'_i(t)$ are independent Gaussian white noise terms for each $i$ with $\langle\eta_i(t)\eta_j(t')\rangle=\Delta_0^2\delta_{ij}\delta(t-t')$ and $\langle\eta'_i(t)\eta'_j(t')\rangle=\Delta_I^2\delta_{ij}\delta(t-t')$. The first term in (\ref{eq:majoranahamiltonian}) is a free-fermion term, whereas the second term is an interaction term. 
Let us define the quantity 
\be
l_{\mathrm{int}} = \f{\Delta_0}{\Delta_I},
\ee
which we will see later is the appropriate crossover lengthscale. 
We will consider the limit of weak interactions, so that $l_{\mathrm{int}}$ is a large parameter.

In this regime, we find that the parameters $\Delta_0$ and $\Delta_I$ in fact drop out of the problem after a rescaling of space and time. That is, the results we obtain for the entanglement line tension are ``universal'' in the sense that they 
apply to the model (\ref{eq:majoranahamiltonian}) with any choice of $\Delta_0$ and $\Delta_I$, so long as $\Delta_I\ll \Delta_0$
(and in fact for a slightly broader class of lattice models which share the same long-wavelength effective description).

We will calculate how two quantities evolve with time: 
the entanglement of the time-evolution operator and the out-of-time-order correlator (OTOC).
These can be expressed using tensor products of the time-evolution operator, ${U(t) = \mathcal{T} \exp \lf i\int \dd t H_\gamma(t) \ri}$, and its complex conjugate.
In order to fix a concrete basis (and therefore a definite convention for complex conjugation),
we may use the Jordan-Wigner  mapping of $U(t)$ to the time-evolution operator of a spin-1/2 chain, with
${i\gamma_{2r-1}\gamma_{2r}\rightarrow X_r}$
and 
${i\gamma_{2r}\gamma_{2r+1}\rightarrow Z_r Z_{r+1}}$
in terms of Pauli matrices.
Choosing the $Z$ basis for the qubits, the operators $i\gamma_i\gamma_{i+1}$ correspond to real matrices. Here we will be schematic: see Ref.~\cite{swann2023spacetimepictureentanglementgeneration} for more detail.

The entanglement of the time-evolution operator \cite{prosen2007operator,
zhou2017operator,
zanardi2001entanglement,
dubail2017entanglement}
is defined by treating the time-evolution operator $\hat U(t)$ as a pure state $\rho^U$ in a doubled Hilbert space $\mathcal{H}\otimes\mathcal{H}$ by an operator-to-state mapping. It is then possible calculate the entanglement of this state. In this paper we focus on the average of the entanglement purity
\begin{equation}\label{eq:S2def}
    e^{-S_2(A)}=\mathrm{Tr}[(\rho^U_A)^2]
\end{equation}
where $\rho_A^U$ is the reduced density matrix of some subsystem $A$ and we have written the purity in terms of the second R\'enyi entropy $S_2(A)$. The ``subsystem'' $A$ can include sites at both the initial time and final time [i.e. both ``input'' and ``output'' legs of $\hat U(t)$, regarded as a tensor network], because these are considered separate sites after the operator-to-state mapping.

The entanglement of the time-evolution operator is the simplest object that gives access to the entanglement line tension \cite{jonay2018coarse}.
Choosing subsystem $A$ to include the region to the left of the origin at the initial time, and to the left of point $x$ at the final time, we expect
\be
S_2(A) \sim \seq \, \mathcal{E}(v) \, t,
 \ee
where $\seq=\f{1}{2}\ln 2$ is the equilibrium entropy density of the physical Majorana chain, 
in units where the lattice spacing is unity, $v=x/t$, and
$\mathcal{E}(v)$ is the ``line tension'' for the membrane.
Above, the constant $\seq$ is
factored out for convenience, in order to simplify some identities obeyed by $\mathcal{E}(v)$ \cite{jonay2018coarse}.
In the present setting it will be useful to further extract
 a factor of a characteristic velocity
$v_B$, which allows 
 $\mathcal{E}$ to be written in terms of a scaling function~$g$ with a dimensionless argument:
\be\label{eq:definescalingformEg}
\mathcal{E}(v) = v_B \, g \lf \f{v}{v_B} \ri.
\ee
To be more precise, the line tension we are computing is that associated with the averaged purity: see Refs.~\cite{ZhouNahum, swann2023spacetimepictureentanglementgeneration} for further discussion of this point.
 
In our 
explicit treatment, $v_B$ will initially appear as a characteristic velocity ${v_c = v_B}$ that emerges from the saddle point equations. However, studying the OTOC shows that $v_B$ may be identified with the ``butterfly velocity'' governing the speed of operator spreading \cite{roberts2015localized,roberts2016lieb,xu2024scrambling}.
With this identification, the line tension that we compute satisfies  conjectured general constraints \cite{jonay2018coarse}.

The OTOC is defined in terms of the expected absolute square of the commutator of two Heisenberg operators of local observables:
\begin{equation}\label{eq:Cdef}
 \mathcal{C} = 
    \frac{1}{2}\mathrm{Tr}\left[\rho_\infty[\hat O_1(0), \hat O_2(t)]^\dagger[\hat O_1(0), \hat O_2(t)]\right]
\end{equation}
where ${\rho_\infty\propto I}$ is the infinite temperature thermal density matrix. If ${t=0}$ and the operators are spatially separated, they will commute and this quantity will be zero. As $t$ increases,  $\hat O_2(t)$ will evolve into an increasingly non-local operator with nontrivial support in some increasingly large region. If $\hat O_1$ is outside of this region, the operators will still approximately commute and the above quantity is still approximately zero, but if $\hat O_1$ is inside this region, they fail to commute and the above quantity is positive and of order 1. $\mathcal{C}$ can therefore be used to measure the spreading of an initially local operator under Heisenberg time evolution. Here these operators will be products of Majorana modes.

Each of  the above observables 
can be written as a matrix element of the ``replicated'' time-evolution operator $U\otimes U^* \otimes U\otimes U^*$; or, after averaging, of
\be\label{eq:replicatedU}
\overline{U\otimes U^* \otimes U\otimes U^*}
=
\exp \lf {-H_\mathrm{eff}t} \ri.
\ee
where the overline represents averaging over the white-noise couplings $\eta$, $\eta'$.
Eq.~\ref{eq:replicatedU} defines the effective Hamiltonian $H_\mathrm{eff}$, which acts on a replicated Hilbert space and will be specified explicitly below.
Note that formally the right-hand side looks like an ``imaginary time'' evolution operator, although the physical dynamics is in real time.

We will show 
that the averaged observables can be approximated by solving coupled classical equations of motion in spacetime which arise from a semiclassical treatment of the path integral for $H_\mathrm{eff}$.
The two observables (\ref{eq:S2def},\ref{eq:Cdef}) differ only in the boundary conditions. 
The resulting classical problem in spacetime involves mobile ``domain walls'' whose initial and final positions are fixed by these boundary conditions, and the problem reduces to calculating the effective action cost of these domain walls as a function of their velocity.

\subsection{Averaging over disorder}

For every different realisation of the random noise $\eta_i(t)$ and $\eta_i'(t)$, the time-evolution operator $U(t)$ will be different, as will the OTOC. 
Prior to noise-averaging, the replicated time-evolution operator ${U\otimes U^*\otimes U \otimes U^*}$
describes four replicas for the Majorana chain evolving simultaneously, with the same realisation of the noise:
\begin{multline} \label{eq:Hrep}
    \Tilde{H}_\gamma(t) =  { \sum_{i,a} } (-1)^{a}[i\eta_i(t)\gamma^a_i\gamma^a_{i+1}+\eta'_i(t)\gamma^a_i\gamma^a_{i+1}\gamma^a_{i+2}\gamma^a_{i+3}].
\end{multline}
We distinguish  the Majorana operators acting on different replicas using the upper index on $\gamma_i^a$, with ${a=1,3}$ for  the copies that evolve with $U$, and ${a=2,4}$ for  the copies that evolve with $U^*$.
Since we are using a basis where $\gamma_i\gamma_{i+1}$ is purely imaginary, 
the complex conjugation for the latter leads to the extra minus signs in (\ref{eq:Hrep}). 
The  noises $\eta$ and $\eta'$ are the the same for all replicas.

We will discuss boundary conditions in detail later. 
As a schematic example, the OTOC is a linear function
of two copies of the thermal density matrix: after the operator-to-state mapping, the two-copies of the thermal density matrix map to the initial pure state (ket) in the replicated system.
The final state (bra) in the replicated system is chosen to yield the desired linear function of the (time-evolved) initial state.

The random variables $\eta$ and $\eta'$ are Gaussian so we can explicitly average over noise to get the effective Hamiltonian $H_\mathrm{eff}=H_0 + H_I$ where
\begin{align}
    H_0 &=
    -\frac{\Delta_0^2}{2}\sum_i
    \left[\sum_a(-1)^{a+1}\gamma^a_i\gamma^a_{i+1}\right]^2,
    \label{eq::H0majorana}\\
    H_I &=
    -\frac{\Delta_I^2}{2}\sum_i
    \left[\sum_a(-1)^{a+1}\gamma^a_i\gamma^a_{i+1}\gamma^a_{i+2}\gamma^a_{i+3}\right]^2.
    \label{eq::HImajorana}
\end{align}
As noted above the evolution of the replicated state, using Eq.~\ref{eq:replicatedU}, resembles what is usually called  imaginary time  evolution.

\subsection{Mapping to spin model}

The effective Hamiltonian above has a simple interpretation as the Hamiltonian of a spin chain (cf. Refs.~\cite{bao2021symmetry,bernard2022dynamics,agarwal2022emergent,agarwal2023charge,swann2023spacetimepictureentanglementgeneration,medvedyeva2016exact}). If we define for each site $i$ the operators
\begin{equation}
    A_i^{ab} \equiv -\frac{i}{2}[\gamma_i^a,\gamma_i^b]
\end{equation}
then the operators $J_i^{ab}\equiv\frac{1}{2}A_i^{ab}$ obey angular momentum commutation relations
\begin{equation}
    [J_i^{ab},J_j^{cd}]=
    i\delta_{ij}(\delta_{ac}J_i^{bd} + \delta_{bd}J_i^{ac}
    - \delta_{ad}J_i^{bc} - \delta_{bc}J_i^{ad})
\end{equation}
These operators form a representation of the Lie algebra $\mathrm{so}(4)$ at each site $i$, so we can think of the operators $A_i^{ab}$ as being the equivalents of Pauli matrices acting on an effective ``spin'' at $i$. Given that there $4$ Majoranas at each site $i$, the local Hilbert space\footnote{As the system is fermionic, the Hilbert space is not strictly a product of local Hilbert spaces, but since the effective model only involves interactions where an even number of Majorana operators act on each site at a time, the 
 dynamics reduces to that of an effective bosonic model.} dimension is $2^{4/2}=4$, and the states in this local Hilbert space transform like spinor representations of $\mathrm{SO}(4)$ 
[or equivalently as representations of $\mathrm{Spin}(4)\simeq \mathrm{SU}(2)\times \mathrm{SU}(2)$]. 
This spinor representation is a direct sum of two 2-dimensional irreducible spinor representations $(1/2,0)$ and $(0,1/2)$. The notation indicates that each irreducible representation transforms nontrivially only under one of the two $\mathrm{SU}(2)$ factors of $\mathrm{Spin}(4)$.

Rewriting  (\ref{eq::H0majorana}) and (\ref{eq::HImajorana}) using these operators gives
\begin{align}
    H_0 &=
    \Delta_0^2\sum_i
    \left[2-\sum_{a<b}A_i^{ab}A_{i+1}^{ab}\right],
    \\
    H_I &=
    \Delta_I^2\sum_i
    \left[2+\sum_{a<b}(-1)^{a+b}A_i^{ab}A_{i+1}^{ab}A_{i+2}^{ab}A_{i+3}^{ab}\right].
\end{align}
The $H_0$ contribution to the effective spin Hamiltonian is just an $\mathrm{SO}(4)$--symmetric ferromagnetic Heisenberg interaction, whereas $H_I$ breaks this continuous symmetry down to a discrete symmetry.
In the replica formalism, this explicit continuous-symmetry-breaking is the key difference between the interacting and the free system. 
It changes the structure of the saddle points and also eliminates the gapless Goldstone modes of the replica theory for the noninteracting system.

For our purposes, the model can be simplified: we can define the chirality operator at site $i$ as $\chi_i=-\gamma_i^1\gamma_i^2\gamma_i^3\gamma_i^4$, which has eigenvalues $\pm 1$, and due to the form of effective Hamiltonian (\ref{eq::H0majorana}) and (\ref{eq::HImajorana}), it follows that each $\chi_i$ is individually conserved. The boundary conditions relevant to entanglement and the OTOC impose the values $\chi_i=+1$ for the conserved quantities on all sites, so we can restrict our Hilbert space to just these states. Spins with $\chi_i=+1$ transform as 2-dimensional irreducible representations of one of the $\mathrm{SU}(2)$ subgroups of $\mathrm{Spin}(4)$.
Therefore the resulting model 
is ultimately quite simple, containing only a single $\mathrm{SU}(2)$  spin-1/2 on each site \cite{bao2021symmetry,agarwal2022emergent,swann2023spacetimepictureentanglementgeneration}, with the Hamiltonian terms
\begin{align}
    H_0 &=
    2\Delta_0^2\sum_i
    \left[1-\Vec{\sigma}_i\cdot\Vec{\sigma}_{i+1}\right] \label{eq:hspin0}
    \\
    H_I &=
    2\Delta_I^2\sum_i
    \left[1+\sum_\alpha(-1)^\alpha\sigma_i^\alpha\sigma_{i+1}^\alpha\sigma_{i+2}^\alpha\sigma_{i+3}^\alpha\right] \label{eq:hspini}
\end{align}
where $\sigma_i^\alpha$ is the the $\alpha$th Pauli for the spin at site $i$ and $\alpha$ is summed over all three Pauli matrices $\alpha=1,2,3$.

In this simplified model, $H_0$ is an $\mathrm{SO}(3)$-symmetric \textit{ferromagnetic} Heisenberg interaction, 
while $H_I$ breaks  $\mathrm{SO}(3)$  down to a discrete $D_4$ symmetry, at the same time opening an energy  gap for local excitations above a ground state. The four ground states are uniform product states of either $\mid\uparrow\rangle$, $\mid\downarrow\rangle$, $\mid\rightarrow\rangle$ or $\mid\leftarrow\rangle$.\footnote{$\mid\uparrow\rangle$, $\mid\downarrow\rangle$ denote $\sigma^3$ eigenstates  and $\mid\rightarrow\rangle$, $\mid\leftarrow\rangle$ denote $\sigma^1$ eigenstates. See Sec.~VII of \cite{swann2023spacetimepictureentanglementgeneration} for further discussion of symmetries of the effective model.}

The problem of computing quantities like entanglement purity or the OTOC as a function of real time $t$ maps to that of calculating  transition amplitudes of the form $\langle \mathrm{final}\vert e^{-\hat H_\mathrm{eff}t}\vert\mathrm{initial}\rangle$, where $\vert\mathrm{initial}\rangle$ and $\vert\mathrm{final}\rangle$ are now states of the $\mathrm{SU}(2)$ spin chain (for both the purity of the time evolution operator and the OTOC, these states are in fact product states). 
In order to calculate such transition amplitudes in the limit of large length and time scales, it will be helpful to use the coherent-state path integral \cite{stone2000semiclassical,tailleur2008mapping}.

\subsection{Coherent state path integral}\label{sec:coherentstate}

Solving for the full imaginary time evolution of the above spin model is intractable. 
However, we expect that we can obtain asymptotically exact results
from the coherent states path integral \cite{stone2000semiclassical,tailleur2008mapping}\cite{swann2023spacetimepictureentanglementgeneration}
when $t$ is large and interactions are weak.

Loosely speaking, this is because 
(\emph{i}) imaginary-time evolution drives  the system towards low-energy-density states at long times, 
and (\emph{ii})~the ground states and low-energy excitations of the model simplify in the limit ${\Delta_I \to 0}$. In that limit, the model is a Heisenberg ferromagnet, 
with ground states that are product states of uniformly aligned spins.
A standard heuristic statement is that in such low-energy regimes, large blocks of almost-aligned spins act like  composite spins with an effective ``large $S$'', so that semiclassics becomes exact.\footnote{See Sec.~VD of Ref.~\cite{swann2023spacetimepictureentanglementgeneration} (published version) 
for a slightly more detailed justification of the exactness of semiclassics which extends to the present case (with $l_\text{int}$ as a large parameter justifying  semiclassics). We will not attempt a  rigorous justification of the exactness of the method in the limit of small $\Delta_I/\Delta_0$ but this would be worthwhile.}

We define single-site coherent states as
\begin{equation} \label{spincoherentstate}
    \vert z_i\rangle = \f{1}{\sqrt{1+z_i^* z_i}}\left(\begin{matrix}1\\z_i\end{matrix}\right),
\end{equation}
and the coherent states of the entire spin chain can be written as products of these states,
${\vert \{z_i\} \rangle =
    \vert z_1 \rangle \vert z_2 \rangle \cdots \vert z_L \rangle}$.
For what follows we drop indices and write $\vert z \rangle \equiv \vert z(x) \rangle$, where the function ${z=z(x)}$ specifies a product state for the entire chain, in continuum notation.

Taking $\Delta_I\to0$ allows us to assume that the spin direction is very slowly varying, and we can take the continuum limit in the  coherent state path integral. Transition amplitudes for the chain take the form
\begin{equation} \label{eq:fullpathintegral}
    \la z_F \vert  e^{-HT} \vert z_I \ra =
    \int_{z(0)=z_I}^{\bar z(T)=z^*_F} 
    \mathcal{D}z \, \exp \lf {-\mathcal{S}} \ri.
\end{equation}
where $T$ is the evolution  time, and where
the action $\mathcal{S}$  is
\begin{equation} \label{eq:actiongeneral}
    \mathcal{S} = \int 
    \dd t \, H(\bar{z},z)  + \mathcal{S}_\mathrm{Berry} + \mathcal{S}_\mathrm{bdry}.
\end{equation}
Here $H(\bar{z},z)=\la z \vert H \vert z \ra$ is  the  ``classical'' Hamiltonian, and we also get a Berry phase term and a boundary term.
(We abuse notation slightly by using the symbol $H$ both for the quantum Hamiltonian and the classical Hamiltonian that appears in the action.)
The  classical Hamiltonian ${H=H_0+H_I}$ is given by\footnote{In taking the continuum limit, we have assumed that $z$ is slowly varying in $x$, since for  $\Delta_I/\Delta_0\to 0$ we expect the gradients to approach zero at sufficiently late times.  Indeed taking the leading terms in a gradient expansion gives a self-consistent theory with gradients $\sim \Delta_I/\Delta_0$. For the quadratic term $H_0$ we take only the lowest term in the gradient expansion (which is quadratic in gradients), and for $H_I$ we take only the zeroth-order term in gradients  (i.e. without derivatives) since $H_I$ already has an explicit factor of $\Delta_I^2/\Delta_0^2$ relative to $H_0$.}
\begin{align} \label{eq:hzero}
    H_0 &= 4\Delta_0^2\int \dd x \, \frac{\bar{z}'z'}{(1+\bar{z}z)^2}, \\
    H_I &= 16\Delta_I^2 \int \dd x \, \frac{\bar{z}z(1-\bar{z}^2)(1-z^2)}{(1+\bar{z}z)^4} \label{eq:hint}.
\end{align}
The Berry phase contribution is
\begin{equation} \label{eq:sberry}
    \mathcal{S}_\mathrm{Berry} =
    -\frac{1}{2}\int \dd x \dd t \,
    \frac{\dot{\bar{z}}z-\bar{z}\dot{z}}{1+\bar{z}z}.
\end{equation}
In computing a transition amplitude, Eq.~\ref{eq:fullpathintegral}, using saddle-point, 
the boundary conditions are as follows.
The field $z$ is fixed only at the \textit{initial} boundary, 
while the field $\bar{z}$ is fixed only at the \textit{final} boundary:
\ba\label{eq:BCzs}
z(x,0)& =z_I(x),
&
 \bar z(x,T)&=z^*_F(x).
\end{align}
The boundary contribution to the action
\begin{equation} \label{eq:boundaryaction}
    \mathcal{S}_\mathrm{bdry} = -\frac{1}{2}\int\mathrm{d}x\ln\frac{(1+z_F^* z(T))(1+\bar{z}(0)z_I)}{(1+z_F^* z_F)(1+z_I^* z_I)}
\end{equation}
depends on $z$ at the final boundary and $\bar{z}$ at the initial boundary, and is zero when $z$ and $\bar{z}$ agree. (Spatial arguments have been suppressed in the above formulas.)

Varying the action $\mathcal{S}$ with respect to $z$ and $\bar{z}$ gives the coupled equations of motion
\begin{align}
    \dot{z} & = - (1+\bar{z}z)^2\frac{\delta H}{\delta\bar{z}}, \\
    \dot{\bar{z}} & = + (1+\bar{z}z)^2\frac{\delta H}{\delta z},
\end{align}
where, crucially, $z$ and  $\bar{z}$ are treated as independent functions of $x$ and $t$: at the saddle point, $\bar z$ need not be the complex conjugate of $z$ \cite{stone2000semiclassical, tailleur2008mapping}\cite{swann2023spacetimepictureentanglementgeneration}.

Explicitly, this gives
\begin{align} \notag
    \dot z =
    & +4\Delta_0^2
    \left(
    z'' -\frac{2\bar z (z')^2}{1+\bar z z}
    \right)\\  \label{eq:eqmotz}
& -16\Delta_I^2
    \frac{z(1-z^2)(1-3\bar{z}^2-3\bar{z}z+\bar{z}^3z)}{(1+\bar z z)^3}, \\ \notag
    \dot{\bar{z}} =
   & -4\Delta_0^2
    \left(
    \bar{z}'' -\frac{2 z (\bar{z}')^2}{1+\bar z z}
    \right)
    \\
    &    +16\Delta_I^2
    \frac{\bar{z}(1-\bar{z}^2)(1-3z^2-3\bar{z}z+\bar{z}z^3)}{(1+\bar z z)^3}.
\end{align}
By solving these equations of motion and substituting the solutions into the action $\mathcal{S}$, we can arrive at a saddle-point approximation for the transition amplitude $\la z_F \vert e^{-Ht} \vert z_I \ra\sim e^{-\mathcal{S}}$.

Given that  $z$ and $\bar{z}$ are both real for the boundary conditions we are interested in, it will be convenient  to use a slightly different parameterisation for much of what follows, with 
\ba\label{eq:angularparam}
z & =\tan\frac{\theta}{2},
& 
\bar{z} & =\tan \frac{\phi}{2}.
\end{align} 
We will switch between using $z$ and $\bar{z}$ and using $\theta$ and $\phi$ as convenient.

The coupled equations of motion for $\theta$ and $\phi$ are
\begin{align}\label{eq:motthetaphi}
    \dot{\theta} &= +4\Delta_0^2 \left(\theta'' + (\theta')^2\tan\frac{\theta-\phi}{2}\right)
    + \Delta_I^2
    F(\theta,\phi), \\ \label{eq:motthetaphi2}
    \dot{\phi} &= -4\Delta_0^2 \left(\phi'' + (\phi')^2\tan\frac{\phi-\theta}{2}\right)
    - \Delta_I^2
    F(\phi,\theta),
\end{align}
where
\be\label{eq:Fdefn}
F(\theta,\phi)=
8\sec^2\frac{{\theta-\phi}}{2} \sin 2\theta
    \left(\tan\frac{\theta-\phi}{2}\sin 2\phi - \cos 2\phi\right).
\ee

It is important to note that  it is possible to set ${\Delta_0=\Delta_I=1}$ 
in the continuum equations of motion
by a rescaling of space and time: 
there are no dimensionless parameters in this continuum problem. 
Dimensional analysis shows that the characteristic length and timescales are  \cite{swann2023spacetimepictureentanglementgeneration}
\ba\label{eq:lengthtimescales}
l & \equiv \f{\Delta_0}{\Delta_I}, 
&
\tau & \equiv \f{1}{\Delta_I^2}.
\end{align}

Since the Hamiltonian is quadratic in the spatial derivatives, the natural generalization of the model would also be rotationally invariant in dimensions higher than one.

\section{Continuum description of the entanglement membrane}
\label{sec:entanglementmembrane}

\begin{figure}
    \centering
\includegraphics[width=\columnwidth]{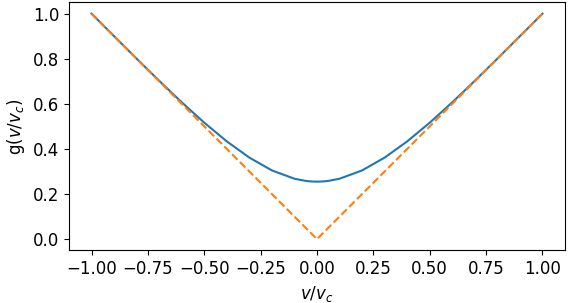}
    \caption{Plot of $\mathcal{E}(v)$ in terms of the dimensionless function $g(v/v_c)$ in blue, calculated numerically. The dotted orange line shows the asymptote $g_\mathrm{asymp}(a)=\vert a\vert$ which $g(v/v_c)$ approaches as $v\to\pm v_c$. (Here $K\equiv\Delta_I/\Delta_0=0.005$ was used in the numerics).} \label{fig:eaginstv}
\end{figure}

The entanglement purity of the time-evolution operator 
can be calculated using a saddle-point approximation by choosing appropriate boundary conditions $z_I$ and $z_F$ and then solving for $z(x,t)$ and $\bar{z}(x,t)$.  As discussed below, the boundary condition $z_I$
imposes a sharp ``domain wall'' \textit{initial} condition for $z$ at the spatial origin, and the boundary condition $z_F$ imposes a domain wall \textit{final} condition in  $\bar z$ at spatial position ${X=v T}$.
In the noninteracting system ($\Delta_I=0$) these domain walls would spread out essentially diffusively  away from the temporal boundaries \cite{swann2023spacetimepictureentanglementgeneration},\footnote{So that at (for example) $t=T/2$ we would have smooth domain walls with a spatial width of order $\sqrt{T}$.}
but we will see that in the interacting system they persist as relatively sharp structures under the  effective time evolution.

We will show  in Sec.~\ref{sec:linetension} that, under certain conditions
(for the present observable, this happens if the the velocity $v$ 
defined by the boundary conditions is not too large)
 the domain walls in $z$ and $\bar{z}$ form a bound state.
By ``bound state'' we simply mean that they remain close together at all times (i.e. the terminology does not imply that the binding has a simple energetic interpretation). The  bound state has an  action cost per-unit-time  $E(v)$ which is a function of velocity $v$, and this bound state is the continuum form of the ``entanglement membrane'' found in other studies of entanglement growth \cite{jonay2018coarse,nahum2017quantum,mezei2018membrane,mezei2020exploring,zhou2020entanglement,rakovszky2019entanglement,agon2021bit,rampp2024entanglement,fisher2023random,brauner2022snowmass}.

In other conditions (such as sufficiently large $v$ for the present boundary conditions), the domain walls  in $z$ and $\bar{z}$ can remain separated  by a parametrically large distance, not forming a bound state but
instead behaving as independent traveling waves each moving at a fixed velocity. We argue that this velocity is the butterfly velocity $v_B$ and that this is also equal to the maximum velocity $v_c$ at which a domain wall bound state can travel  (i.e. $v_c=v_B$). 
We will discuss this unbound case in Sec.~\ref{sec:travellingwaves}.

\begin{figure}
    \centering
\subfigure[]{
\begin{tikzpicture} \label{fig:spboundstate}
    \fill[blue!10] (0,0) rectangle (5,2);
    \draw[thick] (0,0) -- (5,0);
    \draw[thick] (0,2) -- (5,2);
    \draw[thick] (2,0) -- (3,2);
    \fill (2,0) circle (0.08cm);
    \fill (3,2) circle (0.08cm);
    \node at (1,1.2) {$z=0$};
    \node at (1,0.8) {$\bar{z}=0$};
    \node at (4,1.2) {$z=1$};
    \node at (4,0.8) {$\bar{z}=1$};
\end{tikzpicture}
}
\hfill
\subfigure[]{
\begin{tikzpicture} \label{fig:sptravellingwaves}
    \fill[blue!10] (0,0) rectangle (5,2);
    \draw[thick] (0,0) -- (5,0);
    \draw[thick] (0,2) -- (5,2);
    \draw[dashed] (0.5,0) -- (2.5,2);
    \draw[dashed] (2.5,0) -- (4.5,2);
    \fill (0.5,0) circle (0.08cm);
    \fill (4.5,2) circle (0.08cm);
    \draw[dashed,red,thick] (0.5,0) -- (2.5,0);
    \draw[dashed,red,thick] (2.5,2) -- (4.5,2);
    \node at (0.8,1.6) {$z=0$};
    \node at (0.8,1.2) {$\bar{z}=0$};
    \node at (4.2,0.8) {$z=1$};
    \node at (4.2,0.4) {$\bar{z}=1$};
    \node at (2.5,1.2) {$z=1$};
    \node at (2.5,0.8) {$\bar{z}=0$};
\end{tikzpicture}
}
    \caption{Schematic of saddle point solution for the time-evolution operator for different values of $|X/T|$, with space plotted horizontally and time vertically. \ref{fig:spboundstate} shows the solution for $|X/T|<v_c$, with a ballistically travelling bound state connecting the cuts at each boundary (solid line). This is the entanglement membrane. \ref{fig:sptravellingwaves} shows the solution for $|X/T|>v_c$, with travelling waves propagating from each cut with velocity $v_c$ but remaining isolated from each other (dashed lines). The red dashes show the sections of the boundary which disagree and therefore give an extensive contribution to the boundary action.}
    \label{fig:spteo}
\end{figure}
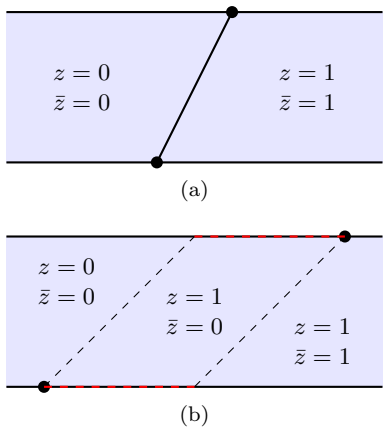

These two situations are shown schematically in Fig.~\ref{fig:spteo}, for the boundary 
conditions relevant to the time-evolution-operator-entanglement. 
(This figure only shows the structure on lengthscales of order $vT$, i.e. it does not resolve the nontrivial structure of the domain walls.)
The boundary conditions in the figure will be described below.

One of the main outcomes is the entanglement line tension.
We anticipate this result in Fig.~\ref{fig:eaginstv}.
The line tension has been written in the form $\mathcal{E}(v) =  v_c g(v/v_c)$ (Eq.~\ref{eq:definescalingformEg}). The function $g$ is then ``universal'' in the sense that all dependence on the microscopic parameters  $\Delta_0$, ${\Delta_I\ll \Delta_0}$, and the lattice spacing, has been absorbed into $v_c$ and $\seq$.
Setting the lattice spacing to unity, 
the critical velocity $v_c$ 
is 
\be\label{eq:vcexactpreview}
v_c =
16\sqrt{2} \, 
\Delta_0 \Delta_I,
\ee
and we will show in Sec.~\ref{sec:otoc} that it may also be identified with the butterfly velocity $v_B$.

Finally, in Sec.~\ref{sec:highvtheory} we will comment on the regime where the speed $|v|$ is very close to, but slightly below, the critical speed $v_c$.
We sketch an analytical argument showing that the spatial extent of the bound state (i.e. of the entanglement membrane) diverges like $1/\sqrt{v_c-|v|}$ as the critical speed is approached, and compare this prediction with numerical results.

\subsection{State entanglement}

We first give a brief review of how the replica model can be used to calculate \emph{state} entanglement,  following a quench \cite{calabrese2004entanglement} from a weakly-entangled state. We will be schematic, as this observable  was covered in detail in \cite{swann2023spacetimepictureentanglementgeneration} for both the free and the interacting case.

In order to calculate the entanglement between a region $A$ of the Majorana chain and its complement $\bar{A}$, we simply choose appropriate boundary conditions for the coherent state path integral.
To find the entanglement of some state of the Majorana chain which has been evolved over a time $T$, we choose the final
state\footnote{In \cite{swann2023spacetimepictureentanglementgeneration}, we reversed the time coordinate $t$ so that this boundary condition was actually applied at $t=0$ and the physical initial state enforced a boundary condition at $t=T$.}  
in the transition amplitude 
 to have all spins pointing right $\mid\rightarrow\ra$ within the region $A$ and all spins pointing up $\mid\uparrow\ra$ outside the region. This fixes the boundary condition for $\bar z$ at $t=T$ (compare Eq.~\ref{eq:BCzs}).

The initial boundary at $t=0$ is instead determined by the physical initial state of the Majorana chain.
 After replication, 
this yields an initial state of the spin chain which is not in fact a coherent state, so slightly more analysis is required. 
(The cases we address below, which are the focus of this paper, will have simpler, coherent state boundary conditions.)
In \cite{swann2023spacetimepictureentanglementgeneration} we argued that generic initial physical states with short-range entanglement enforce a ``reflecting'' boundary condition $z(x,0)=\bar{z}(x,0)$ in the continuum limit.

Let's take the region $A$ to be the right half of the chain $x>0$. The boundary conditions at the final time $t=T$  create a domain wall in $\bar z(x,T)$, with the 
$\ket{\uparrow}$  spins in $\bar A$ imposing ${\bar z(x) = 0}$ for $x<0$ and
the $\ket{\rightarrow}$ spins in $A$ imposing  ${\bar z(x) = 1}$ for $x>0$.

For $\Delta_I=0$ the domain wall  in $\bar z(x)$
relaxes diffusively 
as time is run ``backwards'', away from the final time boundary condition, and in the limit of large the total time $T$ this diffusive spreading allows the domain wall to become arbitrarily wide.
But for ${\Delta_I>0}$, the domain wall eventually relaxes, 
 after a boundary region of temporal extent 
$t_\mathrm{relax}\sim\Delta_I^{-2}$ near the final-time boundary,
to a smooth domain wall with a finite length scale $l_\mathrm{int}\sim\Delta_0/\Delta_I$.

\begin{figure}
    \centering
\includegraphics[width=0.95\columnwidth]{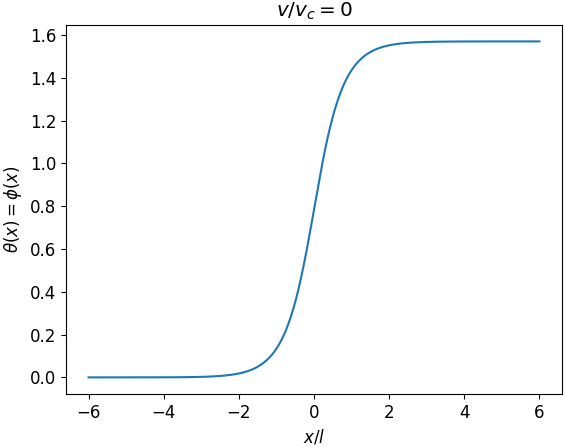}
    \caption{Static solution of the equations of motion with ${\theta(x)=\phi(x)}$, determining the rate of growth of state entanglement after a quench from a weakly-entangled state, and the
    entanglement membrane tension $\mathcal{E}(v)$ at $v=0$. 
    The exact continuum form of the wall is 
       $\theta(x)=\arctan \exp\left(2 x/ l\right)$ 
    (Eq.~\ref{eq:staticsteadystate}). This solution is degenerate under spatial translations. Here $l\equiv \Delta_0/\Delta_I$.}
    \label{fig:staticwall}
\end{figure}

We can find the form of this steady state by noticing that (assuming that, for  $T\gg t_\mathrm{relax}$, a steady state is reached in the region away from the final boundary) the reflecting boundary condition $z(x,0)=\bar{z}(x,0)$ ensures that the steady-state solution satisfies ${z(x)=\bar{z}(x)}$, or ${\theta(x)=\phi(x)}$ in the angular parameterisation of Eq.~\ref{eq:angularparam}. The steady-state property  ${\dot{z}=\dot{\bar{z}}=0}$ means that the Berry phase term is zero. We can then write the spin Hamiltonian as a function of $z$ alone, and the steady-state solution can then be found by simply minimising the classical energy of $z(x)$ subject the constraints that $z\to0$ as $x\to-\infty$ and $z\to1$ as $x\to+\infty$. This gives (switching to the angular parameterisation for convenience)
\begin{equation} \label{eq:staticsteadystate}
    \theta(x)=\arctan \exp\left(2{x}/{l} \right),
\end{equation}
with the 
length scale $l=\Delta_0/\Delta_I$ as defined in Eq.~\ref{eq:lengthtimescales}. 
This solution is shown in Fig.~\ref{fig:staticwall}.
The solution is degenerate under spatial translations, but the symmetry of the boundary conditions ($x\to-x$ and $\theta\to\pi/2-\theta$) implies the above form, with the centre of the smooth domain wall at $x=0$.

\subsection{Entanglement of the time-evolution operator}
\label{sec:opentang}

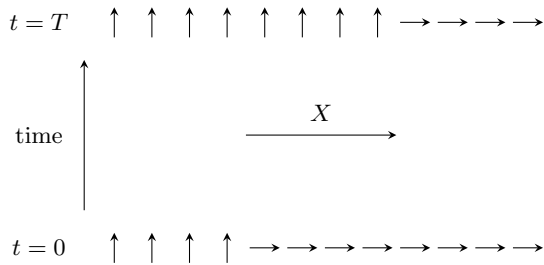
\begin{figure}
    \centering
    \begin{tikzpicture}
        \newcommand{\upspin}[2]{\draw[-stealth] (#1,#2-0.2) -- (#1,#2+0.2);}
        \newcommand{\rightspin}[2]{\draw[-stealth] (#1-0.2,#2) -- (#1+0.2,#2);}
        \upspin{0.5}{0}
        \upspin{1}{0}
        \upspin{1.5}{0}
        \upspin{2}{0}
        \rightspin{2.5}{0}
        \rightspin{3}{0}
        \rightspin{3.5}{0}
        \rightspin{4}{0}
        \rightspin{4.5}{0}
        \rightspin{5}{0}
        \rightspin{5.5}{0}
        \rightspin{6}{0}
        \upspin{0.5}{3}
        \upspin{1}{3}
        \upspin{1.5}{3}
        \upspin{2}{3}
        \upspin{2.5}{3}
        \upspin{3}{3}
        \upspin{3.5}{3}
        \upspin{4}{3}
        \rightspin{4.5}{3}
        \rightspin{5}{3}
        \rightspin{5.5}{3}
        \rightspin{6}{3}
        \node at (-0.5,0) {$t=0$};
        \node at (-0.5,3) {$t=T$};
        \draw[-stealth] (2.25, 1.5) -- (4.25,1.5);
        \node at (3.25, 1.8) {$X$};
        \draw[-stealth] (0.1,0.5) -- (0.1,2.5);
        \node at (-0.5,1.5){$\mathrm{time}$};
    \end{tikzpicture}
    \caption{Initial and final boundary conditions of the spin model used to calculate the entanglement purity of the time-evolution operator. Both boundaries are sharp domain walls with all $\mid\uparrow\rangle$ to left and $\mid\rightarrow\rangle$ to the right, with the position of the domain wall translated by $X$ in final boundary compared to the intial one.}
     \label{fig:boundaryconditions1}
\end{figure}

When calculating state entanglement, the initial boundary of the spin chain is determined by the physical initial state of the Majorana chain. However, we can also calculate the entanglement of the time evolution operator itself \cite{prosen2007operator,
zhou2017operator,
zanardi2001entanglement,
dubail2017entanglement,jonay2018coarse}, 
by treating both initial and final-time surfaces of the time-evolution operator (viewed as a quantum circuit\footnote{Here we briefly use a tensor-network language, which we can most simply visualize by e.g. time-discretizing the evolution operator in the Ising representation (Sec.~\ref{sec:model}) to give a quantum circuit.})
as degrees of freedom of a kind of quantum state whose R\'enyi entanglement entropy can be computed in the usual way. This amounts to performing appropriate contractions for the  initial and final-time ``legs'' (external bonds of the circuit) after replicating  the evolution operator.

Take some subset of these legs to be the subsystem $A$ for which we wish to calculate the entanglement purity. In \cite{swann2023spacetimepictureentanglementgeneration}, we show that ``identity'' index contraction taken outside of the subsystem $A$ corresponds to spins in the state $\mid\uparrow\rangle$, and the ``swap'' index contraction taken within the subsystem $A$ corresponds to spins in the state $\mid\rightarrow\rangle$.
These boundary states are essentially the states, sometimes denoted $\ket{+}$ and $\ket{-}$, that arise in  random circuit calculations from taking traces of powers of the circuit~\cite{fisher2023random}.

To study how the time-evolution operator generates entanglement between different parts of the chain, we will take the subsystem $A$ to be $x>0$ at the initial time $t=0$ and $x>X$ at the final time $t=T$. 
 This yields a transition amplitude for the spin chain (Eq.~\ref{eq:fullpathintegral}) in which  the initial and the final states each  contain a single sharp domain wall, 
with $\mid\uparrow\ra$ to the left of the wall and $\mid\rightarrow\ra$ to its right. 
For $X=0$ these boundary states are identical, but for $X\neq 0$ the boundary states differ by a spatial translation, as shown in Fig. \ref{fig:boundaryconditions1}.

The simplest case is for the operator entanglement is $X=0$, so that both the initial and final boundaries are $\mid\uparrow\ra$ for $x<0$ and $\mid\rightarrow\ra$ for $x>0$. In this case, the problem remains symmetrical under $x\to-x$ (with $\theta\to\pi/2-\theta$ and $\phi\to\pi/2-\phi$)
and the saddle point solution for  $\theta$ and $\phi$ should be roughly constant in time for most of its evolution as long as $T\gg t_\mathrm{relax}$.
The solution will deviate from this steady state significantly only in boundary regions near  $t=0$ and $t=T$.

The symmetries of the problem mean that we should again expect the steady state to satisfy $z=\bar{z}$, and the solution is simply the same steady state (\ref{eq:staticsteadystate}) identified for the state entanglement, with the domain wall staying centred at $x=0$.

\subsection{Moving domain walls and the line tension} \label{sec:linetension}

We now come to the main focus of this paper, which is saddle point solution when boundary conditions force the domain walls to move. The properties of these moving domain walls tell us directly about the entanglement of the time evolution operator, and we will see in Section \ref{sec:otoc} that they also determine the behaviour of the OTOC.

To do this, we consider the case where the cut in the final boundary is translated by some distance $X>0$ relative to the cut in the initial boundary. This more general case contains more information about how information spreads spatially over a distance $X$.

With these boundary conditions, the static domain wall studied in the last section which stays at $x=0$ would incur an extensive boundary action $\propto X$. However, the fact the static is solution is only unique up to translations suggests that there might also be solutions which drift slowly over time, with an additional action-per-time for these solutions that is quadratic in velocity $v$ for sufficiently small $v$. 
Such a drifting domain wall could potentially drift from $x=0$ at $t=0$ to $x=X$ at $t=T$, 
avoiding the boundary cost.

We therefore look for saddle point solutions which move ballistically at constant velocity $v$, i.e. 
\be\label{eq:thetaphimoving}
\theta(x,t)=\Theta(x-vt) \,\,\,\, \text{and} \,\,\,\,  \phi(x,t)=\Phi(x-vt).
\ee
Assuming the solutions have this form, we can convert the problem to a purely spatial one by substituting ${\dot{\theta}=-v\theta'}$ and ${\dot{\phi}=-v\phi'}$, so that Eq.~(\ref{eq:motthetaphi}) becomes
\begin{align}\notag
- v \Theta'(y) = &  
    4\Delta_0^2 \left(\Theta''(y) + \Theta'(y)^2\tan\frac{\Theta(y)-\Phi(y)}{2}\right)
    \\
&    + \Delta_I^2 \,
    F\left( \Theta(y), \Phi(y)\right)\label{eq:steadystategeneral}
\end{align}
where $F$ was defined in Eq.~(\ref{eq:Fdefn}) and 
\be
{y=x-vt}
\ee
is the spatial coordinate in the moving-frame.

Using the symmetry of the problem, we can also  assume that 
\be\label{eq:movingsymm}
\Phi(y)=\pi/2-\Theta(-y).
\ee
We can impose this without loss of generality because the solution
is defined only up to a translation; the point $y=0$ is the ``centre'' of the domain wall by definition.

So finally the saddle-point solution obeys
\begin{align}\notag
- v \Theta'(y) = &  
    4\Delta_0^2 \left(\Theta''(y) + \Theta'(y)^2\tan\frac{\Theta(y)+\Theta(-y)-\f{\pi}{2}}{2}\right)
    \\
&    + \Delta_I^2 \,
    F\left( \Theta(y), \pi/2-\Theta(-y)\right).\label{eq:steadystate}
\end{align}
As noted around Eq.~\ref{eq:lengthtimescales}, the coefficients $\Delta_0$ and $\Delta_I$ can be removed from this equation by rescaling both space and time, implying that the solutions have only a trivial dependence on these parameters.

It seems unlikely that the saddle-point equations can be solved analytically in general, but they can be solved numerically using a ``fictitious dynamics'' that is described in \ref{sec:solvingequations}. 
In short, this method uses the equation of motion (\ref{eq:motthetaphi}) for $\theta(x,t)$ to evolve $\theta$ over time, 
while continuously keeping $\Phi(y)$ equal to  $\pi/2-\Theta(-y)$
as in Eq.~(\ref{eq:movingsymm}).
This evolution converges on the desired steady-state solution of Eq.~\ref{eq:steadystate}.
It is important that  this method only uses the equation of motion for $\theta(x,t)$ 
(while 
continuously enforcing the symmetry which we expect the steady-state solution to have)  --- this avoids the problem that $\phi(x,t)$ has a negative diffusion coefficient in Eq.~\ref{eq:motthetaphi2}. 
The  ``fictitious dynamics'' depends explicitly on $v$ through the symmetry (\ref{eq:movingsymm}) and the fact that $y=x-vt$.

\begin{figure}
    \centering
    \subfigure[]{ \label{fig:movingwall1}
\includegraphics[width=0.95\columnwidth]{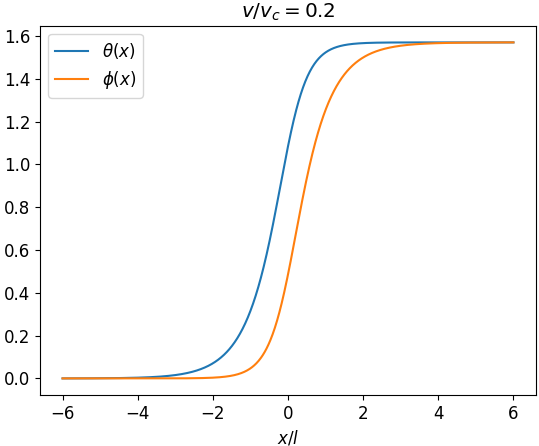}
    }
    \subfigure[]{ \label{fig:movingwall2}
\includegraphics[width=0.95\columnwidth]{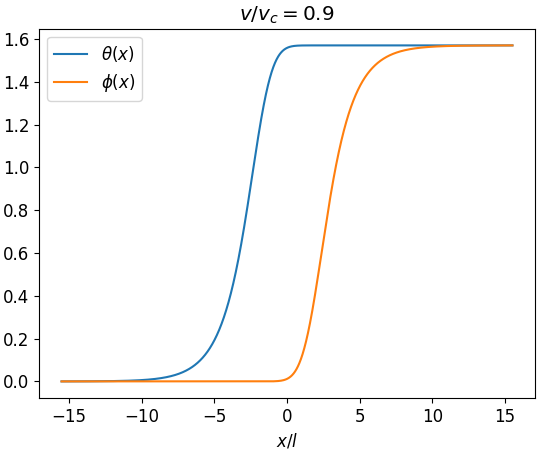}
    }
    \caption{Steady-state solutions of the fictitious dynamics for two velocities in the range $0<v<v_c$. The domain walls are  propagating ballistically to the right with velocity $v$, with $\theta$ ``lagging behind'' $\phi$. Fig.~\ref{fig:movingwall1} shows $v/v_c=0.2$, where the domain walls in $\theta$ and $\phi$ overlap significantly, while \ref{fig:movingwall2} shows $v/v_c=0.9$ where the domain walls propagate almost independently of each other. $x=0$ is the centre of the domain wall which we use to enforce the symmetry between $\theta$ and $\phi$ (Eq.~\ref{eq:movingsymm}). The length scale $l$ is defined as $l\equiv\Delta_0/\Delta_I$. (Here, $\theta(x)$ $\phi(x)$ were calculated numerically with $K\equiv\Delta_I/\Delta_0=0.01$ using the method described in \ref{sec:solvingequations}).}
      \label{fig:movingwall}
\end{figure}

We find numerically that these fictitious dynamics converge to a stable solution whenever the $|v|<v_c$, for the  critical velocity $v_c$ which we will discuss in  the next section. 
We show a couple of examples of stable solutions in Fig.~\ref{fig:movingwall}, 
illustrating that as $v$ increases, 
the domain walls in $\theta$ and $\phi$ pull away from each other.
 (A full discussion of  numerical results is deferred until Sec.~\ref{sec:numerics}.)

Assuming a stable solution can be found for a given $v$, it is straightforward to numerically calculate the action-per-time using (\ref{eq:actiongeneral}). This action has non-zero contributions from both the spin chain Hamiltonian and the Berry phase (the Berry phase is actually purely imaginary, and is therefore a real contribution to the action).

If denote the action per time as a function of $v$ as $E(v)$, 
we can write the purity of the time-evolution operator as $e^{-E(X/T)T+\ldots}$, whenever $|X/T|<v_c$, where the ellipses contain subleading contributions from the regions near the two ends of the domain walls at the temporal boundaries. 
There is no extensive contribution from the boundary, since the length scale $l$ on which the domain wall is smoothed is finite, which means that in Eq.~(\ref{eq:boundaryaction}) $z(T)\approx z_F$ everywhere except within a distance $\sim l$ of the wall.

The dimensional analysis mentioned around Eq.~\ref{eq:lengthtimescales} shows that we can
extract the constants $\Delta_0$ and $\Delta_I$ by writing the cost  (action per unit time) of the bound state in terms of a scaling function $g$ as
\begin{equation}
    E(v) = 
     s_\mathrm{eq} \, v_c \,
    g\left(\frac{v}{v_c}\right) 
\end{equation}
where ${s_\mathrm{eq}=\f{1}{2}\ln 2}$ is the entropy density,
$v_c$ has been defined in Eq.~\ref{eq:vcexactpreview},
and the function $g$ is ``universal'', i.e. independent of $\Delta_0$ and $\Delta_I$.
Note that $E(v)$ differs from $\mathcal{E}(v)$ only by the factor of $\seq$, which is extracted for convenience in  defining  $\mathcal{E}(v)$.

Our numerical result for the line tension, obtained by the method above, is shown in Fig.~\ref{fig:eaginstv}.

Interestingly the numerical result appears consistent with the conjectured general constraints ${\mathcal{E}(v_B)=v_B}$, ${\mathcal{E}'(v_B)=1}$ \cite{jonay2018coarse} 
if we assume that ${v_B=v_c}$. 
In the following Sections we will confirm these identities and show that they arise from nontrivial behaviour of the domain-wall bound state as $|v|\to v_c$.

\subsection{Travelling waves and determination of $v_c$}
\label{sec:travellingwaves}

When $|v|>v_c$, the fictitious dynamics
do not converge numerically to a steady state solution. 
Instead, 
the domain walls in $\theta$ and $\phi$ are widely separated, and the separation grows linearly with time. 
At late times in this fictitious dynamics, the domain wall in $\theta$ is  propagating ballistically in a region where $\phi=0$, but at a speed lower than $v$, so it cannot ``catch up'' with the wall in $\phi$. 

Indeed the true solution for ${v>v_c}$ 
involves well-separated domain walls in $\theta$ and $\phi$. We  first study the behaviour of the travelling wave in $\theta$ (in a region where $\phi=0$) on its own. 
 
It is actually more convenient to return to the parameterization using $z$ instead of $\theta$ for this problem. 
The reader is reminded that $z=\tan\frac{\theta}{2}$ and in particular
\begin{center}
    \begin{tabular}{>{\centering\arraybackslash}m{2em}|>{\centering\arraybackslash}m{2em} >{\centering\arraybackslash}m{2em} >{\centering\arraybackslash}m{2em} >{\centering\arraybackslash}m{2em}}
        $\theta$ & $-\frac{\pi}{2}$ & $0$ & $\frac{\pi}{2}$ & $\pi$ \\ [0.5ex]
        \hline
        $z$ & $-1$ & $0$ & $1$ & $\infty$
    \end{tabular}
\end{center}
so the initial boundary conditions are ${z=0}$ for ${x<0}$ and ${z=1}$ for ${x>0}$.

The equation we get for $z$ when we set $\bar{z}=0$ is a travelling wave equation of KPP-Fisher type  \cite{kolmogorov1study,fisher1937wave}:
\begin{equation} \label{eq:travellingwave}
    \dot{z} = 4\Delta_0^2 \, z'' - 16\Delta_I^2 \, z (1 - z^2).
\end{equation}
After completing this work we learned that this equation has also appeared in recent work on the multi-flavour version of the Majorana chain \cite{parrikar2025entanglement}. 
The formalism there is somewhat different, but the approach is also based on a semiclassical treatment of the replica theory.

\begin{figure}
    \centering
\includegraphics[width=0.95\columnwidth]{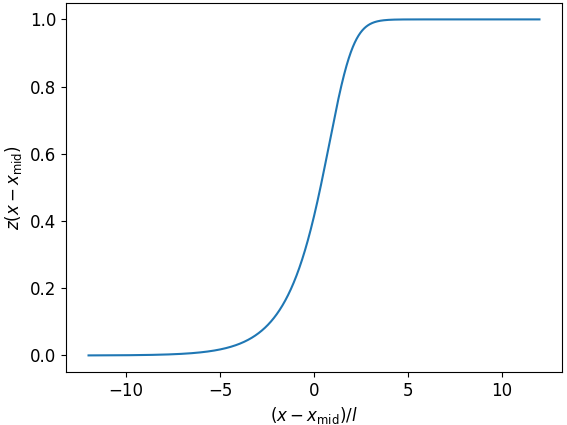}
    \caption{An example of a right-moving travelling wave solution for $z$. The midpoint $x_\mathrm{mid}$ is chosen arbitrarily as the point where $\theta=2\arctan z=\pi/4$. (Here, $z(x)$ was found numerically with $K\equiv\Delta_I/\Delta_0=0.01$).}
    \label{fig:travellingwave}
\end{figure}

This equation admits ballistically travelling solutions at all speeds. However, for a wide range of intial conditions (including the sharp domain wall in our case), the steady state solution travels at a fixed velocity $v_*$ which can be calculated analytically to be 
\be\label{eq:travwavespeed}
v_* = 16\sqrt{2}\Delta_0\Delta_I
\ee
(see Appendix \ref{app:fisherkpp} for details). This solution is shown in Fig.~\ref{fig:travellingwave}.

Having understood a single domain wall (in $z$), now let us consider what it tells us about the full problem, with domain walls in both $z$ and $\bar z$. 
Recall that a velocity ${v=X/T}$ is defined  by the boundary conditions 
for computing the operator entanglement, and that the
``fictitious dynamics''\footnote{See Sec.~\ref{sec:linetension} for a brief description of the fictitious dynamics  and Sec.~\ref{sec:solvingequations} for more detail.}
is a method we can use to search for bound-state solutions moving at a given speed $v$.

First consider the low-$v$ regime. We imagine  running the fictitious dynamics with ${0<v<v_*}$, and with the domain walls for $z$ and $\bar z$ initially well-separated, with the domain wall in $z$ to the left.
Then the $z$-domain wall will move as the travelling wave discussed immediately above, at speed $v_*$ in the rest frame.
Such a right-moving travelling wave in $z$ (or $\theta$)
will eventually catch up the domain wall in $\bar z$
(which is its mirror image in the frame moving at speed $v$)
and collide with it. 
One may argue that the domain wall in $z$ cannot \textit{overtake} the domain wall in $\bar{z}$  (or a contradiction would result).\footnote{Because then it would enter a region where $\bar z\approx 1$, while $z$ still obeys the boundary conditions $z\to0$ as $x\to-\infty$ and $z\to1$ as $x\to+\infty$. Under a change of basis which sends $z=0$ to $z=1$ and vice versa, this would again give us Eq. (\ref{eq:travellingwave}), but with opposite boundary conditions on $z$ i.e.
$z\to1$ as $x\to-\infty$ and $z\to0$ as $x\to+\infty$. This would result in a travelling wave moving to the \emph{left} with velocity $v_*$, which is clearly a contradiction.} If the domain wall in $z$ cannot lag indefinitely behind the domain wall in $\bar z$, and it also cannot overtake it, the two domain walls must form a bound state travelling with velocity $v$.

Therefore for $v<v*$ we indeed expect the fictitious dynamics to converge to a true bound-state solution of the saddle-point equations.

On the other hand if $|v|>v_*$, the domain wall in $\theta$ never catches up with its mirror image and travels only as an isolated domain wall, described by (\ref{eq:travellingwave}).
Similarly, when we turn from the fictitious dynamics to to the true boundary conditions (Sec.~\ref{sec:opentang}) 
then when $v>v_*$
there is a plausible picture [summarized below and in Fig.~\ref{fig:sptravellingwaves} above] with  well-separated domain walls in $z$ and $\bar z$ that travel along parallel worldlines at speed $v_*$.

So whenever $|v|<v_*$ we expect a bound state to form, and whenever $|v|>v_*$, we expect no bound state. In other words, we identify $v_*$ with the critical velocity $v_c$ for domain-wall unbinding that was discussed above.
 This identification fixes 
the exact value of  $v_c$ quoted above in Eq.~\ref{eq:vcexactpreview},
in agreement with the numerics.

The complete saddle point solution for $|v|>v_c$ has an isolated domain wall in $\theta$ travelling from $x=0$ at $t=0$ at velocity $v_*$, reaching the final boundary at $x=v_*T$. Similarly, an isolated domain wall in $\phi$ starts at $x=X$ at the final time $t=T$ and travels backwards in time at velocity $v_*$, reaching the initial boundary at $x=X-v_*T$. 

As for the bound state regime, we can calculate the action per unit time in the ${v>v_c}$ regime. Here, the Hamiltonian contribution is zero so the action per time comes entirely from the Berry phase, which we can calculate exactly as $\frac{1}{2}\ln 2 \cdot v_*$ (we get $\frac{1}{4}\ln 2 \cdot v_*$ for each wall). However, unlike for the bound state, there is also a significant boundary contribution, because on each boundary there is a parametrically large gap between the domain walls, where $\theta$ and $\phi$ do not agree. The boundary term gives $\frac{1}{2}\ln 2$ on every site on the boundary for which the two fields do not match, so $\frac{1}{2}\ln 2\cdot (X-v_* T)$ in total.

The total action is therefore simply $\mathcal{S}=\frac{1}{2}\ln 2 \cdot X$ and the entanglement purity is $\sim e^{-\frac{1}{2}\ln 2 \cdot X}$  for ${v>v_c}$.

We see that there is a qualitative difference in the entanglement of the time-evolution operator as a function of $X$ and $T$ according to whether we are inside or outside the ``light cone'' given by  ${|X/T|= v_c}$. This qualitative change is associated with the existence of a bound state of the domain walls in $z$ and $\bar{z}$ only for ${v<v_c}$.

\subsection{Behaviour of the bound state as $|v|\to v_c$} \label{sec:highvtheory}

We can use the insights of the previous section to investigate the behaviour of the bound state solutions (i.e. the steady-state solutions of Eq.~\ref{eq:steadystate} for $|v|<v_c$) as $|v|\to v_c$ from below.
In this limit, the domain walls in $\theta$ and $\phi$ become further and further apart, and their influence on each other diminishes, and we expect them to increasingly resemble independent travelling waves.

In terms of $z$ and $\bar{z}$, if $\bar{z}$ really were zero everywhere, then $z$ would be described exactly by Eq.~(\ref{eq:travellingwave}), and it would have a travelling wave solution travelling at exactly $v_c$. The fact that $\bar{z}$ is not exactly zero, and is in fact $\approx1$ in some region sufficiently far in front of the domain wall in $z$, must lead to the domain wall in $z$  travelling with a lower velocity $v<v_c$.

Ref.~\cite{brunetderrida} analysed the effect of imposing a cutoff of size $\varepsilon$ 
at the front of a travelling wave 
(a value of the variable below which the exponential growth term is set to zero), 
and found that this significantly reduces the velocity of the travelling wave, by an amount $\propto 1/(\ln \varepsilon)^2$ at small $\varepsilon$.
They also argued that this modification to the velocity was quite robust to the specific way in which the cutoff is implemented.
The basic mechanism involves the matching of solutions to the linearized equation for the front in the regions where the cutoff is/is not active. This approach was later used to study phase transitions in travelling wave problems induced either by a wall moving at an imposed velocity  \cite{derrida2007survival} or by a  region that the travelling wave is inhibited from entering \cite{nahum2021measurement} (related to population growth in a shrinking domain, and to the recursive treatment of tree tensor networks, respectively).

In our case, we do not have a cutoff for $1-z$ imposed in advance, or a velocity imposed in advance.
Nevertheless we expect that a similar mechanism applies.
Effectively, in the equation for $z$  we have a position-dependent cutoff that is determined by the form of $\bar z$. From (\ref{eq:eqmotz}), the linearised equation for ${\delta z = 1 - z}$ reads
\be
 \delta \dot z = 4 \Delta_0^2 \delta z'' + 32 \Delta_I^2 \left[ 1 - \f{6 \bar z(x)}{\lf 1+\bar z(x)\ri^2} \right] \delta z.
\ee
In the region where $\bar z\approx 0$, we get a travelling wave equation like Eq.~(\ref{eq:travellingwave}) which is unstable around $\delta z=0$, whereas in the region where $\bar z\approx 1$, we get a travelling wave equation which is \emph{stable} around $\delta z=0$. This leads to a more rapid decay of the solution at large $x$ than would be obtained for Eq.~(\ref{eq:travellingwave}) and is effectively like a cutoff on the $\delta z$ equation with $\varepsilon$ equal to whatever value $\delta z$ has at the position of the domain wall in~$\bar{z}$. 
Matching solutions to the left and the right of this cutoff \cite{brunetderrida} gives a relation between 
the velocity and the separation of the domain walls. 
Of course we could equally well apply the argument to the equation for $\bar z$, instead of that for $\delta z$, and we should get a consistent result: here that is guaranteed by symmetry.

\begin{figure} 
    \centering
    \includegraphics[width=0.95\columnwidth]{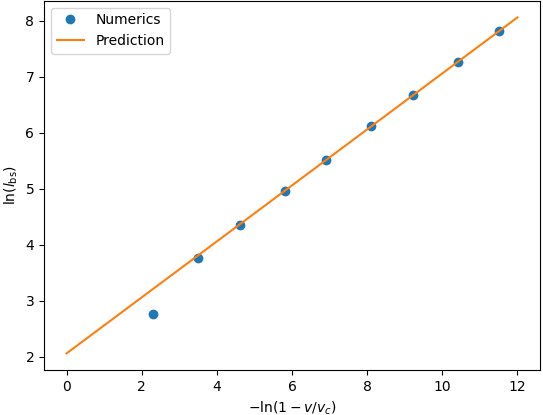}
    \caption{A log-log plot of the size of the bound state $l_\mathrm{bs}$ against $1-v/v_c$ in the limit as $v\to v_c$, comparing numerical results from the iterative method against the prediction Eq.~(\ref{eq:boundstatedivergence}). Given that $l_\mathrm{bs}$ is only defined up to an additive constant, here the additive constant is chosen to given best agreement with the predicted asymptotic behaviour.  (That is, the brown line includes one fitting parameter.) Here $\Delta_I/\Delta_0=0.1$.}
    \label{fig:highvlbs}
\end{figure}

\begin{figure}
    \centering
    \includegraphics[width=0.95\columnwidth]{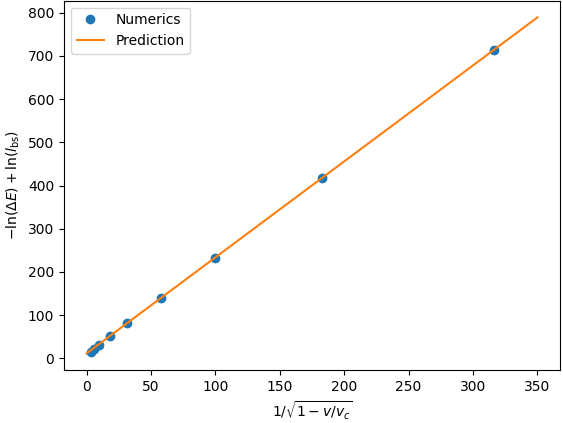}
    \caption{Numerical test of the prediction in Eq.~(\ref{eq:eathighvprediction}), using the iterative method. We take the log of both sides of the equation and group the $\ln(\Delta E)$ and $\ln(l_\mathrm{bs})$ terms together for clarity. 
    In addition to the constant in $l_{\rm bs}$ that was fixed in Fig.~\ref{fig:highvlbs}, the brown line in the present figure includes one new fitting parameter, which is a vertical offset, corresponding to  the undetermined overall constant in Eq.~\ref{eq:eathighvprediction}.
In the numerics the parameterisation ${1-z = e^{-h}}$ is used to resolve small values of ${1-z}$.
(Here ${\Delta_I/\Delta_0=0.1}$.)}
     \label{fig:highvdeltaE}
\end{figure}

Taking the distance between the domain walls in $z$ and $\bar{z}$ to be $l_\mathrm{bs}$, the matching argument gives \cite{brunetderrida}
\begin{equation}
    k_\mathrm{im} l_\mathrm{bs} \sim \frac{\pi}{4}
\end{equation}
where $k_\mathrm{im}$ is the imaginary part of the wavenumber $k$ in the solution to the linearised equation $1-z\propto e^{-k(x-vt)}$, with $k_\mathrm{im}$ being non-zero for $|v|<v_c$. We can easily calculate $k_\mathrm{im}=l^{-1} \sqrt{1-v/v_c}$ which gives a prediction for $l_\mathrm{bs}$ in the limit $|v|\to v_c$,
\begin{equation} \label{eq:boundstatedivergence}
    l_\mathrm{bs} \sim \frac{\pi}{4}
    \frac{l}{\sqrt{1-v/v_c}}.
\end{equation}
This prediction is compared with numerical simulations  in  Fig.~\ref{fig:highvlbs}.
These numerical results are
obtained using an iterative procedure, described in Sec.~\ref{sec:highvnumerics}, which efficiently finds self-consistent solutions for $|v|\approx v_c$.

We can also use the above approximation to predict the behaviour of the line tension $\mathcal{E}(v)$ as $|v|\to v_c$. Assuming the fronts of the domain walls resemble the fronts of the travelling waves, and therefore decay exponentially at a rate $k\approx2\sqrt{2}\Delta_I\Delta_0$
as you move into the interior of the region between the two domain walls, it follows that ${\bar z (1-z)}$ is of order 
 $e^{-kl_\mathrm{bs}}$ in
this interior region. Since the nontrivial part of the action is $O\lf \bar z (1-z)\ri$, a naive guess for the correction to the bound state cost 
(relative to the cost of independent domain walls in $\bar z$, $z$) is\footnote{Note that the order of limits we are considering here is ${\Delta_I/\Delta_0\to 0}$ first, and then $v\to v_c$.}
\begin{equation} \label{eq:eathighvprediction}
    \Delta E(v)=E(v)-\frac{1}{2}\ln 2\cdot |v|\propto
    l_\mathrm{bs}\exp\left(-\frac{\pi}{\sqrt{2}}\frac{1}{\sqrt{1-|v|/v_c}}\right)
\end{equation}
as $|v|\to v_c$.
This prediction is compared with numerics in Fig.~\ref{fig:highvdeltaE}.

\section{Out-of-time-order correlator} \label{sec:otoc}

Now we will see how saddle-point solutions closely related to those above can be used to compute the out-of-time-order correlator (OTOC), which is a tool for measuring how Heisenberg operators spread under time evolution \cite{roberts2015localized,roberts2016lieb,xu2024scrambling}. 

Take a local operator $\hat A$ supported on some patch  near the origin. 
It commutes with all local operators on  sites outside the patch, but not generally with local operators on the same patch. 
Under Heisenberg time evolution, the operator becomes a linear combination 
of products of Majoranas with larger and larger support, spreading in all directions. Therefore, the time-evolved operator $\hat A(T)$ has a larger commutator with local operators in an ever increasing volume of space. We can quantify this by calculating the expected square of the commutator of the time-evolved operator $\hat A(T)$ and a local operator $\hat B(0)$ at position $X$
\begin{equation}
\mathcal{C}(X,T) = 
    \frac{1}{2}\Tr\left[\rho_\infty[\hat A(T),\hat B(0)]^\dagger[\hat A(T),\hat B(0)]\right],
\end{equation}
where $\rho_\infty$ is the infinite temperature density matrix, which is the equilibrium state of the noisy Majorana chain. We will also assume that $\hat A=\hat A(0)$ and $\hat B =\hat B(0)$ are bosonic operators, so that they correspond to local observables. This means that they are products of \emph{even} numbers of Majorana operators.

If we take $\hat A$ and $\hat B$ to be products of Majorana operators (with appropriate factors of $i$ so that they are Hermitian), then $\hat {A}^\dagger=\hat A$, $\hat {B}^\dagger=\hat B$ 
and $\hat {A}^2=\hat{B}^2=1$, which gives us the simplified expression
\begin{equation}
 \mathcal{C}(X,T) = 
    1 - \frac{1}{\mathcal{N}}\Tr[\hat A(T)\hat B(0) \hat A(T) \hat B(0)]
\end{equation}
where $\mathcal{N}$ is the dimension of the Hilbert space. The second term in the expression is the OTOC.

If the operator $\hat B$ which acts locally at position $X$ is far outside the typical support of the Majorana strings appearing in the expansion of $\hat A(T)$, then $\hat A(T)$ and $\hat B$ essentially commute and the OTOC should be $\approx 1$. However, as the strings appearing in $\hat A(T)$ grow to include $X$, the OTOC should drop to a smaller value (potentially to zero). At any time $T$, the OTOC should therefore have a lower value within some region, and be $\approx 1$ outside this region, with this region growing over time. If this region grows ballistically over time, the velocity with which it grows is called the \emph{butterfly velocity} $v_B$ \cite{roberts2015localized,roberts2016lieb,aleiner2016microscopic,von_Keyserlingk_2018,Nahum_2018}. This velocity measures the speed at which operators spread out in space under Heisenberg time evolution.

In this Section we first describe how to compute the OTOC and $v_B$ using the coherent-state path integral, and in Sec.~\ref{sec:markov}
 we compare with an alternative calculation
 using an effective Markov process for strings of Majorana modes.
 This is analogous to the Markov picture for the OTOC in random circuits \cite{Nahum_2018,von_Keyserlingk_2018}.
 
We will discuss only the leading-order semiclassical result, which gives $v_B$ in the limit of small $\Delta_I/\Delta_0$. We leave corrections to this semiclassical result (which may be nontrivial), as well as crossovers close to the lightcone,
to the future.
In the present regime, the Markov picture leads to a particularly simple traveling wave equation. 

An interesting range of different travelling wave equations have appeared in  studies of the OTOC, including 
in Fermi liquids
\cite{aleiner2016microscopic,aron2023traveling,aron2023kinetics},
long-range-interacting systems \cite{zhou2025operator,chen2019quantum,zhou2023hydrodynamic},
and quantum circuits in a mean-field-like  regime \cite{Nahum_2018,xu2019locality} or with weak interactions \cite{keselman2020scrambling,delucaANunpublished}.
In particular, Ref.~\cite{agarwal2022emergent}
 derived a travelling wave equation, coupled to charge transport, for the OTOC in a noisy fermion chain with U(1) charge conservation symmetry.

\subsection{OTOC boundary conditions}

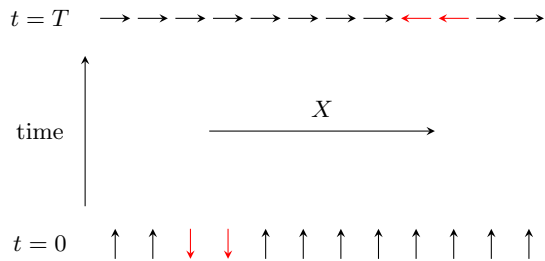
\begin{figure}
    \centering
    \begin{tikzpicture}
        \newcommand{\upspin}[2]{\draw[-stealth] (#1,#2-0.2) -- (#1,#2+0.2);}
        \newcommand{\rightspin}[2]{\draw[-stealth] (#1-0.2,#2) -- (#1+0.2,#2);}
        \newcommand{\downspin}[2]{\draw[-stealth] (#1,#2+0.2) -- (#1,#2-0.2);}
        \newcommand{\leftspin}[2]{\draw[-stealth] (#1+0.2,#2) -- (#1-0.2,#2);}
        \upspin{0.5}{0}
        \upspin{1}{0}
        {\red\downspin{1.5}{0}
        \downspin{2}{0}}
        \upspin{2.5}{0}
        \upspin{3}{0}
        \upspin{3.5}{0}
        \upspin{4}{0}
        \upspin{4.5}{0}
        \upspin{5}{0}
        \upspin{5.5}{0}
        \upspin{6}{0}
        \rightspin{0.5}{3}
        \rightspin{1}{3}
        \rightspin{1.5}{3}
        \rightspin{2}{3}
        \rightspin{2.5}{3}
        \rightspin{3}{3}
        \rightspin{3.5}{3}
        \rightspin{4}{3}
        {\red\leftspin{4.5}{3}
        \leftspin{5}{3}}
        \rightspin{5.5}{3}
        \rightspin{6}{3}
        \node at (-0.5,0) {$t=0$};
        \node at (-0.5,3) {$t=T$};
        \draw[-stealth] (1.75, 1.5) -- (4.75,1.5);
        \node at (3.25, 1.8) {$X$};
        \draw[-stealth] (0.1,0.5) -- (0.1,2.5);
        \node at (-0.5,1.5){$\mathrm{time}$};
    \end{tikzpicture}
    \caption{Initial and final boundary conditions of the spin model used to calculate the averaged OTOC, where both $\hat A$ and $\hat B$ are of the form $-i\hat\gamma_i\hat\gamma_{i+1}$. The initial boundary $t=0$ is $\mid\uparrow\rangle$ everywhere except where the Majorana operators in $\hat A$ act, where it is $\mid\downarrow\rangle$, and the final boundary $t=T$ is $\mid\rightarrow\rangle$ everywhere except where  the Majorana operators in $\hat B$ act, where it is $\mid\leftarrow\rangle$.}
     \label{fig:boundaryconditions}
\end{figure}

We can use the same approach that we used to calculate the average purity of the time-evolution operator to OTOC in an infinite temperature  state (since the noisy chain has no extensive conserved quantities, this is the natural choice of state).
To do this, we simply choose the appropriate boundary conditions. Let's say both of the local operators are simply products of neighbouring Majorana operators $\hat A=-i\hat\gamma_i\hat\gamma_{i+1}$ and $\hat B=-i\hat\gamma_j\hat\gamma_{j+1}$. In this case, the initial boundary is all up apart from at $i$ and $i+1$, where the spins are down, $\mid\cdots\uparrow\uparrow\downarrow\downarrow\uparrow\uparrow\cdots\ra$, and the final boundary is all right apart from at $j$ and $j+1$ where the spins are left, $\mid\cdots\rightarrow\rightarrow\leftarrow\leftarrow\rightarrow\rightarrow\cdots\ra$: see Fig.~\ref{fig:boundaryconditions}.

\subsection{Travelling wave solutions for ${|X/T|>v_*}$}

The saddle point solution with the above boundary conditions is a bit less obvious than it was for the entanglement of the time-evolution operator. Let us first consider the simpler case in which  $X$ is very large, i.e. $X\gg v_*T$. Let's focus on what happens near the origin. The initial boundary of $z$ is all up except for one spin at the origin which is down. The final boundary is all right. We choose a basis such that right corresponds to $z=0$ and up corresponds to $z=1$.

In this basis and in this limit, $\bar{z}=0$ everywhere  near the origin, so we can apply the travelling wave equation (\ref{eq:travellingwave}),
which we repeat for convenience:
\begin{equation} \label{eq:travellingwaverepeat}
    \dot{z} = 4\Delta_0^2 z'' - 16\Delta_I^2 z (1 - z^2).
\end{equation}
The initial condition for $z$ is $z=1$ everywhere except at the origin where $z=-1$. With these initial conditions, we get two ballistically propagating travelling waves, one moving in each direction away from the origin. In front of these waves $z=1$, and in their wake $z=0$ (not $z=-1$, because this is unstable; the stable state is $z=0$). Over time, these travelling waves relax to the same travelling wave solutions as before (Sec.~\ref{sec:travellingwaves}).

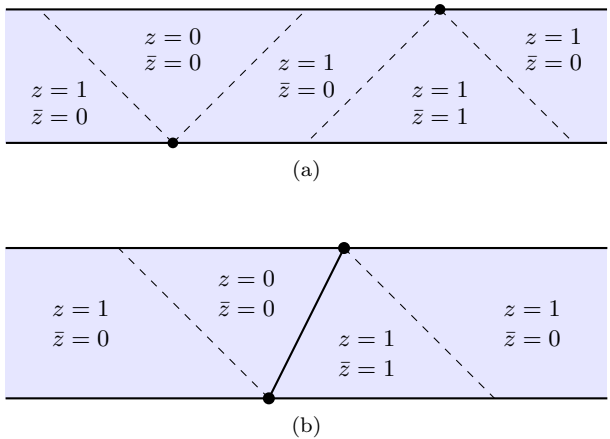
\begin{figure}
    \centering
\subfigure[]{
\begin{tikzpicture}[scale=0.888]
\label{fig:spotochighv}
    \fill[blue!10] (0,0) rectangle (9,2);
    \draw[thick] (0,0) -- (9,0);
    \draw[thick] (0,2) -- (9,2);
    \draw[dashed] (2.5,0) -- (0.5,2);
    \draw[dashed] (2.5,0) -- (4.5,2);
    \draw[dashed] (6.5,2) -- (4.5,0);
    \draw[dashed] (6.5,2) -- (8.5,0);
    \fill (2.5,0) circle (0.08cm);
    \fill (6.5,2) circle (0.08cm);
    \node at (2.5,1.6) {$z=0$};
    \node at (2.5,1.2) {$\bar{z}=0$};
    \node at (6.5,0.8) {$z=1$};
    \node at (6.5,0.4) {$\bar{z}=1$};
    \node at (4.5,1.2) {$z=1$};
    \node at (4.5,0.8) {$\bar{z}=0$};
    \node at (0.8,0.8) {$z=1$};
    \node at (0.8,0.4) {$\bar{z}=0$};
    \node at (8.2,1.6) {$z=1$};
    \node at (8.2,1.2) {$\bar{z}=0$};
\end{tikzpicture}
}

\bigskip

\subfigure[]{
\begin{tikzpicture} \label{fig:otoclowv}
    \fill[blue!10] (0,0) rectangle (8,2);
    \draw[thick] (0,0) -- (8,0);
    \draw[thick] (0,2) -- (8,2);
    \draw[thick] (3.5,0) -- (4.5,2);
    \draw[dashed] (3.5,0) -- (1.5,2);
    \draw[dashed] (4.5,2) -- (6.5,0);
    \fill (3.5,0) circle (0.08cm);
    \fill (4.5,2) circle (0.08cm);
    \node at (1,1.2) {$z=1$};
    \node at (1,0.8) {$\bar{z}=0$};
    \node at (7,1.2) {$z=1$};
    \node at (7,0.8) {$\bar{z}=0$};
    \node at (3.2,1.6) {$z=0$};
    \node at (3.2,1.2) {$\bar{z}=0$};
    \node at (4.8,0.8) {$z=1$};
    \node at (4.8,0.4) {$\bar{z}=1$};
\end{tikzpicture}
}
    \caption{Schematic of saddle point solution for the OTOC for different values of $|X/T|$. \ref{fig:spotochighv} shows the solution for $|X/T|>v_c$, where a pair of travelling waves propagate from each boundary point at velocity $v_c$, without meeting. \ref{fig:otoclowv} shows the solution for $|X/T|<v_c$, where two of these travelling waves have coalesced into a bound state, while the other two travelling waves are unaffected.}
    \label{fig:spotoc}
\end{figure}

If we assume that this is the solution, and that,  by symmetry, the same thing happens in reverse  (``backwards in time'') for $\bar{z}$ in the region around $x=X$, then we can calculate the OTOC. We get a contribution from the Berry phase term of $\ln 2 v_* T$ and a contribution from the boundary of $-\ln 2 v_* T$. These terms cancel out exactly to give $\mathcal{S}=0$ and the OTOC is 1. This is  indeed what we should expect when $X$ is sufficiently large. This solution should hold whenever $|X/T|>v_*$ as $T\to\infty$, 
 since $|X/T|>v_*$ ensures that the  travelling waves in $z$ and $\bar z$ remain well-separated.

\subsection{Bound state solutions}

Let's now consider what happens when ${|X/T|<v_*}$ with ${X>0}$.

If we were to continuously decrease $X$ from above $v_*T$ to below, the right-moving travelling wave in $z$ would collide with the one in $\bar{z}$. 
The natural guess is that 
this reduces the speed of both travelling waves, so that they travel together at speed ${v=X/T}$, 
in the bound state solution
that we found in the context of the  time-evolution operator entanglement: see Fig.~\ref{fig:spotoc}~(b).
The two outer travelling waves, moving in the other direction, are not affected. 

The complete solution for $X>0$ therefore resembles Fig.~\ref{fig:otoclowv}, 
 with ``outer'' traveling waves in $z$ (on the left) and $\bar z$ (on the right),
and a bound state moving at velocity $v=X/T$ that connects ${x=0,t=0}$ with ${x=X,t=T}$.

Once again we can calculate the action. We get contributions from the travelling waves, the bound state, and the boundaries. Putting it all together we get a positive action
\begin{equation} \label{eq:otocaction}
    \mathcal{S} = \left(E\left(\frac{X}{T}\right)-\frac{1}{2}\ln 2 \cdot\left\vert\frac{X}{T}\right\vert\right)\cdot T,
\end{equation}
 where  $E(X/T)T=\seq \mathcal{E}(X/T)T$ is the nontrivial contribution from the bound state.

This action is proportional to $T$ 
for any fixed value of value of $X/T$, 
and therefore the OTOC is exponentially small, $\sim e^{-\mathcal{S}}$. 
We see therefore that while the OTOC is 1 outside the light cone $X/T=v_*$, it approaches zero everywhere within this light cone as $T\to\infty$. 
We therefore identify the velocity $v_*$ with the butterfly velocity $v_B$.
This identification will be confirmed via the alternative calculation below (Sec.~\ref{sec:markov}).

By symmetry we obtain an analogous picture for ${X<0}$. Interestingly the resulting solution is not smoothly connected to the one we have discussed for ${X>0}$.
At first sight it is surprising that the saddle point solution shown in Fig.~\ref{fig:otoclowv} should change discontinuously as $X$ changes from positive to negative in order to remain symmetric in $X$, given that there is no obvious change in the OTOC at this point. 
However, both saddle point solutions (one with left-moving travelling waves and one with right-moving travelling waves) actually exist for both positive and negative $X$, although for $X\neq0$ the contribution of one of these will be exponentially smaller in $|X|$ than the other. 
This has a close analogue in the random unitary circuit, where the OTOC is related to the probability that the point $(X,T)$ lies  in between  the endpoints two biased random walks associated with the left and right boundaries of the spreading operator. 
This can fail to be true either because the right-moving walk ends up to the left of $(X,T)$, or because the left-moving walk ends up to the right of $(X,T)$, but for $|X|\gg 1$, one of these possibilities is exponentially less likely than the other.

\subsection{Independent computation of $v_B$:  \\ Effective Markov process}
\label{sec:markov}

The ballistic spreading of the OTOC can also be understood in terms of a Markov process. Let operator $\hat A(t)$ be expressed as sum of products of Majorana operators
\begin{equation}\label{eq:opexpansion}
    \hat A(t) = \sum_S a_S(t)\hat\Gamma_S
\end{equation}
where $S$ runs over all subsets of sites containing an even number of sites\footnote{If we assume that $\hat A$ is a bosonic operator which is initially a product of Majorana operators, it must initially be a product of an even number of Majorana operators. Under local time evolution, the operator must remain bosonic, and therefore must be linear combination of products of even numbers of Majorana operators.},
and $\hat\Gamma_S$ is the product over all Majorana operators in $S$,
\begin{equation}\label{eq:Gammadef}
    \hat\Gamma_S=(-i)^{|S|/2}\prod_{i\in S}\hat\gamma_i,
\end{equation}
where the ordering of the product is such that smaller values of $i$ appear to the left of larger values. The operator $\hat A(t)$ is then fully described by the real coefficients $a_S(t)$. Note that $\Tr[\hat\Gamma_S\hat\Gamma_{S'}]=\mathcal{N}\delta_{SS'}$, so $\Tr[\hat A(t)^2]=\mathcal{N}\sum_Sa_S(t)^2$ is constant over time. If we choose $\hat A(0)$ to be a product of Majorana operators then $\sum_Sa_S(t)^2=1$ at all $t$.

In Appendix \ref{app:markov} we show that the noise-averaged values $\overline{a_S(t)^2}$ obey a master equation corresponding to a fictitious Markov process on subsets of sites (clusters)~$S$,
\begin{equation}\label{eq:markovschematic}
    \frac{d}{dt}\overline{a_{S}(t)^2}=\sum_{S'} R_{SS'}\overline{a_{S'}(t)^2},
\end{equation}
where
$\overline{a_{S}(t)^2}$ is interpreted as the time-dependent  probability  of configuration $S$ in the fictitious Markov process and
$R_{SS'}$ describes the transition rate from configuration $S'$ to configuration $S$.

The fictitious Markov process involves ``particles'' at the sites $i\in S$ which hop to unoccupied neighbouring sites at a rate $4\Delta_0^2$; the number of particles can also change when four contiguous sites which contain an odd number of particles all have their occupancies reversed (so one particles becomes three over vice versa) which happens at a rate $4\Delta_I^2$.

The limit $\Delta_I^2\ll\Delta_0^2$ corresponds to the limit where the particles diffuse for long periods of time and only split or recombine very occasionally. The fact that $\Delta_I^2\ll\Delta_0^2$ is a continuum limit in the spin model seems to correspond to the fact that, when the particles are sparse, we can think of them as being continuous random walks with diffusion coefficient $4\Delta_0^2$ which only occasionally split into three walks at a rate $4\Delta_I^2$
 (and only occasionally recombine).

Define the density of particles $\rho_i$ as the the probability that site $i$ contains  a particle, where the probability of a string is  $\overline{a_S(t)^2}$: i.e. $\rho_i=\sum_{S|i\in S}\overline{a_S(t)^2}$. If we ignore fluctuations and assume that the particles locally equilibrate much faster than they  split or recombine, we can get a continuum equation for the density
\begin{equation}
    \frac{\partial \rho}{\partial t}
    =4 \Delta_0^2 \frac{\partial^2 \rho}{\partial x^2}
    + 32\Delta_I^2\rho(1-\rho)(1-2\rho).
    \label{eq:otoctravwave}
\end{equation}
This is again a Fisher-KPP-like equation: it is actually the same equation as (\ref{eq:travellingwave}) if we make the substitution $\rho=(1-z)/2$. 
It admits a travelling wave solution with velocity $v_*=16\sqrt{2}\Delta_0\Delta_I$.

The speed of this travelling wave is the growth rate of the (right-hand boundary of the) operator support, so we again obtain $v_B=v_*$, consistently with the saddle-point approach to the replica spin model.

The Markov mapping in Eq.~\ref{eq:markovschematic}
is exact, while 
the travelling wave equation (\ref{eq:otoctravwave}) arises from a mean-field approximation to this Markov process which we expect to be valid in the limit of small $\Delta_I/\Delta_0$. To go beyond the mean-field approximation we should account for the discreteness of the effective ``particles''. 
Similar effects have been discussed in the context of Brownian quantum circuits in Ref.~\cite{xu2019locality}:
a general theory for the effect of discreteness on travelling wave fronts \cite{brunetderrida,brunet2001effect}  shows that the subleading corrections to $v_B$ are nontrivial and nonanalytic in $\Delta_I/\Delta_0$.

\section{Numerics} \label{sec:numerics}

In Sec.~\ref{sec:entanglementmembrane} we compared analytical results on the entanglement membrane tension to numerics. We now give more information on how these simulations were performed.

The equilibrium equation (\ref{eq:steadystate}) for a domain wall bound state travelling with velocity $v$ is analytically intractable, but can be solved approximately through numerical integration of the ``fictitious dynamics'' mentioned in Sec.~\ref{sec:linetension}.
We detail this approach in
Sec.~\ref{sec:solvingequations}.

However, as $v\to v_c$ from below, the fictitious dyanmics take arbitrarily long to converge to a steady state solution, so it becomes impractical to study the limiting behaviour of the bound state and $\mathcal{E}(v)$ using this method.

Instead, we use a different iterative method that involves first guessing the form of $\bar{z}$ (which approaches a travelling wave as $v\to v_c$) and then solving for $z$ by using the known behaviour of $z(x)$ as $x\to+\infty$. We can then iterate this procedure by using the resulting solution of $z$ to form our new guess for $\bar{z}$. This procedure converges to a self-consistent solution very quickly when $v\approx v_c$. We discuss this alternative method in Sec~\ref{sec:highvnumerics}.

\subsection{Numerically solving the equilibrium equations} \label{sec:solvingequations}

How do we find  moving bound state   solutions for $\theta(x,t)$ and $\phi(x,t)$ in practise? One way is to use the equations of motion.

To motivate the approach, let us first imagine a hypothetical situation where we are \textit{given} 
the form of $\phi$ 
(up to a translation) in the steady state, and only have to determine $\theta$.

As in Eq.~\ref{eq:thetaphimoving}, we move to the moving frame coordinates $(y,t)$, writing ${\Theta(y,t) = \theta(y+vt, t)}$ etc. 
Then the supplied data is a fixed function $\Phi(y)$ in the moving frame.

Working in this  frame,
we could evolve $\Theta(y,t)$ forwards in time, 
starting from an initial state   $\Theta(y,0)$ with a sharp domain wall 
[$\Theta(y)=0$ for $x<0$ and $\Theta(y)=\pi/2$ for $x>0$], and using the steady-state form for the other field: $\Phi(y,t)=\Phi(y)$.
The equation of motion for $\Theta$ is ``diffusive'' in the sense that $\dot{\Theta}\sim+\Theta''+\cdots$,
with a positive coefficient for the $\Theta''$ term 
(contrast the equation for $\Phi$, where this term has the opposite sign).
We might therefore expect that as $t\to\infty$,  $\Theta(y,t)$ 
will approach the desired solution  $\Theta(y)$  to the equilibrium equation (\ref{eq:steadystategeneral}), assuming this solution does exist.

In practice, we do not know the solution for $\Phi(y)$ in advance.
We cannot find the equilibrium solution by evolving both
$\Theta(y,t)$ and $\Phi(y,t)$ forward in time together, because 
 the equation of motion (\ref{eq:motthetaphi2}) is  $\dot{\Phi}\sim-\Phi''+\cdots$ with a \emph{minus} sign, meaning that $\Phi$ only evolves diffusively if we evolve \emph{backwards} in time
(evolving forward in time with a generic initial condition will cause small devations from the equilibrium solution to \emph{grow} exponentially over time).

Fortunately, we do not need to evolve $\Phi$ independently from $\Theta$ because we already know that reflection symmetry imposes
${\Phi(y)=\pi/2-\Theta(-y)}$ (\ref{eq:movingsymm}) in the steady state. Therefore we introduce a fictitious dynamics in which we evolve $\Theta$ forwards in time using its equation of motion, and continuously update $\Phi$ so that it satisfies the reflection symmetry of the expected steady state solution.
If this dynamics converges to a steady state, then this is automatically a solution of Eq.~(\ref{eq:steadystate}).
Explicitly,  the fictitious dynamics is 
\begin{align}\notag
& \dot \Theta( y,t)  =  v \Theta'(y,t) \\  \notag
& + 4\Delta_0^2 \left(\Theta''(y,t) + \Theta'(y,t)^2\tan\frac{\Theta(y,t)+\Theta(-y,t)-\f{\pi}{2}}{2}\right) \\
& + \Delta_I^2 \,
    F\left( \Theta(y,t), \pi/2-\Theta(-y,t)\right),
    \label{eq:fictitious}
\end{align}
where the primes are derivatives with respect to $y$ and $F$ is defined in Eq.~\ref{eq:Fdefn}. 
Since $\Theta$  has a positive diffusion coefficient,
 the dynamics is stable. 
If $\Theta$ approaches a steady state, then this is a saddle-point solution which travels at velocity $v$. 

The way the velocity $v$ enters the problem is in how the ``centre'' of the domain wall moves (the point around which we reflect $\theta$ to get $\phi$).  
 This leads to the explicit $v$-dependence in Eq.~(\ref{eq:fictitious}).
Note that we can try to force the centre to move at any given $v$, but we are not guaranteed a priori that a steady state solution (satisfying the boundary conditions as $y\to \pm \infty$) will exist for all $v$.

Numerical integration of Eq.~(\ref{eq:fictitious}) shows that for sufficiently low $|v|<v_c$,
with 
\be\label{eq:vcnumerical}
v_c \simeq 22.6\cdot \Delta_0 \Delta_I
\ee
there is a stable ``bound state'' solution, with domain walls in $\theta$ and $\phi$ propagating together, whose action per unit time is a function of $v$. 
 (The above estimate for $v_c$ is consistent with the analytical result, see Eq.~\ref{eq:vcexactpreview} and Sec.~\ref{sec:travellingwaves}.)
However, for $|v|\geq v_c$, there is no stable solution.

\subsubsection{Implementation} \label{sec:fictitiousdynamics}

In order to simulate the fictitious dynamics described above, we first discretise space. In principle, there is a natural discretisation which comes from using the equations of motion for the underlying discrete spin model (\ref{eq:hspin0}) and (\ref{eq:hspini}), but for simplicity we do not use this. Instead, we simply make the substitution $\theta_i'\to{(\theta_{i+1}-\theta_{i-1})/2}$ and $\theta_i''\to{\theta_{i+1}+\theta_{i-1}-2\theta_i}$, taking the lattice spacing to be one.

Initially taking $\theta_i=0$ for $i<0$ and $\theta_i=\pi/2$ for $i>0$ (and $\theta_0=\pi/4$) we update $\theta_i$ in each time step according to
\begin{align} \notag
    \theta_i\to
    \theta_i+[
    v\theta_i'
    &+4\theta_i''
    +4(\theta_i')^2\tan\frac{\theta_i+\theta_{-i}-\pi/2}{2} \\
    &+K^2 F\left(\theta_i,\pi/2-\theta_{-i}\right)
    ] \delta t \label{eq:numericalupdate}
\end{align}
where $K$ sets the length scale $l\equiv K^{-1}$. We can think of this as setting $\Delta_0=1$ and  $\Delta_I=K$, where we take the limit of small $K$.
The parameter $\delta t$ denotes the size of the small time step taken, but given that we are looking for steady state solutions where the bracketed term in (\ref{eq:numericalupdate}) approaches zero, the results should be insensitive to this parameter, and it is only important that it is small enough for numerics to be stable.

To determine whether or not $\theta_i$ has converged, it is important to find a measure of how quickly $\theta_i$ is changing which scales appropriately with $K$ (smaller values of ${K=l^{-1}}$ will lead to smaller derivatives and therefore slower convergence). 

Clearly the total change $\sum_i|\dot\theta_i|$ will contain a factor of $l$ because the sum will include more sites with large values of $|\dot\theta_i|$, so this can be corrected with a factor of $l^{-1}=K$. However, as mentioned above, we also expect smaller values of $K$ to converge over longer time scales.
We need to distinguish between the case where $\sum_i|\dot\theta_i|$
 is small because the solution has converged, and the case where it is small just because the dynamics is slow.

Take two functions, $f_1(t)=ce^{-k_1 t}$ and $f_2(t)=ce^{-k_2 t}$ which agree at $t=0$ but which decay exponentially at different rates. The condition $f_1(t_1)=f_2(t_2)$ that they both decayed by the same amount translates into the condition on the derivatives $t_1\dot f_1(t_1)=t_2\dot f_2(t_2)$.

We  therefore use the following measure for convergence
\begin{equation}
    c_K(t) = Kt\sum_i|\dot\theta_i(t)|
\end{equation}
and  run the numerics (\ref{eq:numericalupdate}) until $c_K(t)$ is decreasing and is below some fixed threshold. For the results in this paper, we used a threshold of $10^{-4}$, which gives quantitative results very close to those achieved with smaller thresholds, as long as $|v|/v_c$ is not too close to $1$ (for $|v|/v_c$ close to $1$ we used a different method described in the following section).

There are two non-trivial checks on the validity of the numerical approach: firstly, when ${v=0}$ it should agree with the known analytic solution ${\theta(x)=\arctan\exp(2Kx)}$; secondly, the numerical value for the velocity $v_c$ above which the bound state gives way to travelling waves should agree with the analytic prediction $v_c=16\sqrt{2}K$.

Taking $v=0$, the numeric approximation accurately reproduces the predicted functional and gives the following predicted values for $E(0)$
%
% 0.02      - 0.03999644597270544
% 0.01      - 0.01999955955965932
% 0.005     - 0.009999952607318275
% 0.0025    - 0.005000011389768192
\begin{center}
    \begin{tabular}{>{\centering\arraybackslash}m{4em}|>{\centering\arraybackslash}m{4em} >{\centering\arraybackslash}m{6em} >{\centering\arraybackslash}m{6em}}
        $K$ & $E(0)$ predicted & $E(0)$ numerics & ratio \\ [0.5ex]
        \hline
        0.02 & 0.04 & 0.03999645... & 0.999911... \\
        0.01 & 0.02 & 0.01999956... & 0.999978... \\
        0.005 & 0.01 & 0.00999995... & 0.999995...
    \end{tabular}
\end{center}
showing good agreement.

With regards to a qualitative change at $|v|=v_c$, there is clear convergence when $v$ is significantly less than $v_c$, and clear divergence when $v$ is significantly larger than $v_c$, 
 but isolating $v_c$ numerically on this basis is difficult because the time taken to converge becomes very large as $v\to v_c$.

\subsection{Alternative method when $v\approx v_c$} \label{sec:highvnumerics}

Since the previous method becomes impractical as $|v| \to v_c$, we study this limit using a completely different method, which converges \emph{faster} for $|v|$ close to $v_c$.

The basic idea of the alternative method is  to make an initial guess $\bar z_0(y)$ for $\bar z(y)$, and solve the  --- purely spatial --- stationary equation for $z(y)$, obtaining  an approximation $z_1(y)$. Using the reflection symmetry, this gives us an improved estimate $\bar z_1(y)$ for $\bar z(y)$. 
The process can therefore be iterated to obtain a sequence of estimates $(\bar z_k(y), z_k(y))$ which we hope will converge to the true solution.

When $v\lesssim v_c$ this method is effective because we have a good initial estimate $\bar z_0(x)$.  This initial estimate is given by the free travelling wave solution from Sec.~\ref{sec:travellingwaves}. 
As discussed in Sec.~\ref{sec:highvtheory}, the  individual wavefronts making up the bound state are close to the isolated travelling wave form (up to the reflection that relates $z$ and $\bar z$) when $v_c-v$ is small.

In each step we have to solve the spatial equation which has two derivatives and therefore requires two boundary conditions.
Fortunately, we can reduce the two unknowns to a single one by solving the equation in the limit ${x\to+\infty}$. In this limit, the equation of motion for $z$ can be linearised around $z=1$, giving the form of $z$ up to a constant factor. This allows us the fix the logarithmic derivative $z'/z$ at the boundary $x\to+\infty$.
This still leaves one unknown e.g. the value of $z(x_0)$ at some point $x_0$ far to the right, which we deal with using the shooting method.\footnote{We guess $z(x_0)$. If the value we guessed is too high, then when we continue the solution for $z$ to the left using the equation \ref{eq:steadystategeneral}, $z$ will diverge in the positive direction. If the value we guessed was too small, it will diverge in the negative direction. As a result, we can always tell whether the value we guessed was too large or too small. Once we have an upper bound $z_\mathrm{max}$ and lower bound $z_\mathrm{min}$ on the true value, we can check the middle of this interval $(z_\mathrm{max}+z_\mathrm{min})/2$ and improve exactly one of these bounds, halving the window. By iterating this procedure, we essentially perform a binary search, finding the true value to arbitrary accuracy in logarithmic time.}

We find that, for $|v|$ sufficiently close to $v_c$, the solution for $z$ converges after only a few iterations, giving an efficient numerical method to find self-consistent solutions as $v\to v_c$.
We  used this method to obtain the data in Figs.~\ref{fig:highvlbs},~\ref{fig:highvdeltaE} above, where we set
${\Delta_I/\Delta_0 = 0.1}$. For $|v|/v_c\geq0.9$, the solution essentially converges after a single iteration.

In the limit $|v|\to v_c$, the forms of domain walls approach those of free travelling waves, meaning that the very first solution we find for $z$ should already be very close to the self-consistent one i.e. even without iterating.

\section{Conclusions and outlook}

By mapping the $n=2$ replica model of the weakly interacting Majorana chain to a spin chain, and rewriting it as a coherent-state path integral, 
we get a continuum model for operator entanglement which can be solved using a saddle point approximation. The equations of motion involve two fields, $z(x,t)$ and $\bar{z}(x,t)$, 
or equivalently $\theta(x,t)$ and $\phi(x,t)$, arising initially as parameterisations of spin directions.

If the boundary conditions enforce  domain walls, the domain walls in $\theta$ and $\phi$ can form a bound state, which can travel at a range of velocities $|v|<v_c$ for a critical velocity $v_c$. 
These bound state solutions acquire an action per time $E(v)=\seq \mathcal{E}(v)$ and correspond to the entanglement membrane found in previous studies, with $\mathcal{E}(v)$ being the line tension of the membrane for $|v|<v_c$.

If the boundary conditions are such that a bound state would need to travel at ${|v|>v_c}$, a bound state does not form, and instead the domain walls in $\theta$ and $\phi$ 
travel independently at the critical velocity $v_c$, each being described by a  Fisher-KPP-like equation.
The analysis also shows how the size of the bound state (entanglement membrane) diverges as the critical speed is approached,~$l_\mathrm{bs}\sim1/\sqrt{1-|v|/v_c}$.

The line tension $\mathcal{E}(v)$ and the qualitative change at ${|v|=v_c}$ allow for predictions about the entanglement of the time-evolution operator and the averaged OTOC. In turn this allows us to identify the critical velocity $v_c$ with the butterfly velocity $v_B$.

An advantage of the effective continuum model is a weak form of  universality of the results: a range of noisy Majorana lattice models, with weak four-fermion interactions, would show the same effective continuum action 
(\ref{eq:actiongeneral}) with some effective parameters $\Delta_0$ and  $\Delta_I$.
In turn these parameters can be eliminated by rescaling space and time, so that the line tension is governed by a ``universal'' scaling function  $g(v/v_c)$. 
This means that  on large length and time scales, the scrambling properties we have discussed depend only on a length scale $l\equiv \Delta_0/\Delta_I$, a time scale $\tau\equiv\Delta_I^{-2}$ and the equilibrium entropy density~$s_\mathrm{eq}$.

The length and time scales $l$ and $\tau$ 
mark the crossover between free fermion and interacting behaviour; formally this is due to a crosssover from continuous to discrete symmetry \cite{swann2023spacetimepictureentanglementgeneration,nahum2021measurement} which changes the nature of  the low-lying excitations of the effective model and  its saddle-point solutions.

Now let us mention  possible extensions and some questions that we have left open.

First, we have looked here only at two observables, but the coherent-states formalism  introduced here and in Ref.~\cite{swann2023spacetimepictureentanglementgeneration} could be applied to a much wider range of observables that are expressible using ${n\leq 2}$ replicas, such as  moments of correlation functions
\cite{nahum2022real,yoshimura2025operator,bernard2019open,
bernard2021solution,
bernard2021can,
bernard2022dynamics,
hruza2024qssep,
barraquand2025introduction},
frame potentials (extending the extensive study in random circuits, see e.g.
\cite{hunter2019unitary,schuster2025random}), or cross-entropies \cite{ware2023sharp,morvan2024phase}.

It would also be interesting to go beyond leading order in the semiclassical expansion. For example, the Markov process picture  can be used to understand subleading corrections to $v_B$ \cite{xu2019locality,brunetderrida} that are nonanalytic in $v_I/v_0$, and it would be interesting to see how this effect arises in the path integral approach.

The formalism could also be applied to systems with conserved quantities, as in the noisy QSSEP model \cite{bernard2019open,
bernard2021solution,
bernard2021can,
bernard2022dynamics,
hruza2024qssep,
barraquand2025introduction,albert2026universal}, at the cost of additional fields (compare the monitored systems in \cite{poboiko2023theory,fava2024monitored}), see forthcoming work \cite{scopainprogress}.

For us it was the limit   of weak interactions that enabled  exact, or numerically exact, results.
This limit did not prevent us from studying 
the genuinely interacting features of quantum chaos: as noted above it simply means that we must examine observables beyond the large crossover scale $l$ in order to see the interacting ``universality class''.
Similarly, if we move away from the limit of small $\Delta_I/\Delta_0$, we anticipate that  the basic bound state picture will remain intact, 
with the  line tension $\mathcal{E}(v)$ receiving additional corrections that are subleading at small $\Delta_I/\Delta_0$ and which may be computable by going beyond the leading order in semiclassics.

However, it is not immediately obvious to what extent our results for the regime ${v\approx v_c}$ are sensitive to the limit of weak interactions that we took (recall that we took this limit prior to approaching $v_c$). What happens as ${v\to v_c}$ at fixed interaction strength?  We conjecture that there will still be a diverging bound state size, and an unbound regime for ${|v|>v_c}$.

For this and for other questions, it may be enlightening to make a more detailed comparison between the random circuit formalism 
(where at $N=2$ we usually work with a single ``Ising-like'' variable, essentially related to $z$) and the present formalism with two continuum fields that are (loosely speaking) related by time-reversal.

Another open question concerns the fluctuations in (for example) the second R\'enyi entropy \cite{ZhouNahum}. 
For the free model, we found numerically in Ref.~\cite{swann2023spacetimepictureentanglementgeneration} that fluctuations were sufficiently weak that averaging the purity and averaging the entropy gave the same asymptotic forms at long times: $\overline{S_2} = -\ln \overline{e^{-S_2}} + \text{subleading}$.\footnote{This equality can be checked analytically at very late times, after saturation. In the monitored free Majorana  system the equality of the two averages (``annealed'' and ``quenched'') at leading order follows fairly simply from the  structure of saddle points in the replica formalism~\cite{sigmamodelmeasurement}.}
One might speculate that, here, the two kinds of averages give the same result for $\mathcal{E}(v)$ in the limit of small $\Delta_I/\Delta_0$. One way to approach this is by considering a more general number of replicas and an appropriate limit.

Another challenge that may in principle be addressable  in the  continuum formalism using the replica trick is to calculate the  von Neumann entropy, rather than the second R\'enyi entropy.

\smallskip

\textbf{Note added:}  While we were completing this manuscript we learned of related work by M. Mezei and H. Rajgadia \cite{mezei2025entanglement} which also derives the structure of the entanglement membrane in noisy Majorana chains (Brownian SYK chains). While the approaches used are quite different, the results appear to be consistent where they overlap.

\acknowledgments 
 We thank John Chalker, Denis Bernard, Fabian Essler and Andrea De Luca for useful
 discussion and encouragement and for collaboration on previous related work. 
 We thank the authors of Ref.~\cite{mezei2025entanglement} for correspondence.
TS was supported by a James Buckee Scholarship and by Royal Society Enhancement Award (RGF$\backslash$EA$\backslash$181042).
AN is supported by the European Union (ERC, STAQQ, 101171399). Views and opinions expressed are however those of the authors only and do not necessarily reflect those of the European Union or the European Research Council Executive Agency. Neither the European Union nor the granting authority can be held responsible for them.

\appendix

\section{Angular parameterisation}

The free-fermion Hamiltonian density is
\begin{equation}
    h_0 = \Delta_0^2\sec^2\frac{\theta-\phi}{2}\theta'\phi'
\end{equation}
and the interaction Hamiltonian density is
\begin{equation}
    h_I = \Delta_I^2 \sec^4\frac{\theta-\phi}{2}
    \sin 2\theta \sin 2\phi
\end{equation}

The Berry phase term can be rewritten using
\begin{align}
    \frac{\dot{\bar{z}}z-\bar{z}\dot{z}}{1+\bar{z}z}
    &= \frac{\dot{\theta}+\dot{\phi}}{2}\tan\frac{\theta-\phi}{2}
    - \frac{\dot\theta}{2}\tan\frac{\theta}{2}
    + \frac{\dot\phi}{2}\tan\frac{\phi}{2} \\
    &= \frac{\dot{\theta}+\dot{\phi}}{2}\tan\frac{\theta-\phi}{2}
    + \frac{\partial}{\partial t}
    \left [
    \ln \cos \frac{\theta}{2} - \ln \cos \frac{\phi}{2}
    \right]
\end{align}

\section{Boundary conditions for the entanglement of the time evolution operator} \label{app:toe-bcs}

In order to calculate the entanglement purity of a pure quantum state of the Majorana chain, it is necessary to take a particular ``index contraction'' of the replicated state, which involves taking the inner product of the states of different replicas which are paired up in a particular way.

For example, the ``identity'' contraction corresponds to taking the inner product of the state of each ket replica with its corresponding bra replica (so $a=1$ with $a=2$ and $a=3$ with $a=4$).

Another important index contraction for $N=2$ is the ``swap'' contraction, where we take the inner product of the state of each ket replica with \emph{other} bra replica (so $a=1$ with $a=4$ and $a=3$ with $a=2$).

To calculate the purity of a subsystem $A$, the identity contraction should be applied outside $A$ to take the partial trace with respect to $A$ and give $\rho_A\otimes\rho_A$, and the swap contraction should be applied within $A$ to give $\Tr[\rho_A^2]$.

In \cite{swann2023spacetimepictureentanglementgeneration} we show that taking the identity contraction at a given spatial site corresponds to contracting with 
$\vert\up\ra$ at the final time, setting a final-time boundary condition at that site, 
while the swap contraction corresponds to the boundary condition $\mid\rightarrow\ra$. 
Calculating the purity within a region $A$ for a time-evolved pure state $\hat U(t)\ket{\psi}$ therefore corresponds to fixing the final state $\mid\rightarrow\ra$ within $A$ and $\vert\up\ra$ elsewhere.

For this pure state entanglement problem we perform the index contractions only at the final time.
However, to calculate the entanglement purity of the time evolution operator $\hat U(t)$ itself, we should make the index contractions at both the initial time and final time, so that in the replica model, the replicated initial state $\ket{\psi}\otimes\ket{\psi}^*\otimes\ket{\psi}\otimes\ket{\psi}^*$ of the pure state problem is also replaced by a state corresponding to an index contraction.

Clearly then, in the $SO(3)$ Heisenberg chain, both the initial  and final boundary conditions  involve contracting with $\mid\rightarrow\ra$ within some region $A$ (where $A$ can now include portions of both the initial and final time surfaces) and $\vert\up\ra$ outside of $A$.

\section{Travelling wave review} \label{app:fisherkpp}

 Eq.~\ref{eq:travellingwave}, which we repeat for convenience,
\begin{equation} 
\label{eq:travellingwaverepeat}
    \dot{z} = 4\Delta_0^2 z'' - 16\Delta_I^2 z (1 - z^2),
\end{equation}
is a Fisher-KPP type equation \cite{kolmogorov1study,fisher1937wave,bramson1983convergence,
derrida1988polymers,
brunet2015exactly,munier2014lecture}
which can be thought of as the continuum limit of a chain of population growth models with random hoppings between neighbours.
 Let us briefly review some well-known features.

The equation has two relevant fixed points, a stable one with $z=0$ everywhere and an unstable one with $z=1$ everywhere. 
If initially $0\leq z(x)\leq 1$ for all $x$, then this constraint will be true at all times. For generic initial conditions satisfying this constraint, 
 $z(x)$ will at every $x$ eventually approach the stable value $z=0$, 
but this will take a long time if $x$  is initially inside a large region with $z=1$. 

Consider the case where $z(x)$ is initially a step function with $z(x)=0$ for $x<0$ and $z(x)=1$ for $x>0$. Then at late times the solution is a travelling wave, with a $z\sim 0$  region to the left of the wavefront invading a $z\sim 1$ region to its right,
at an asymptotically  constant velocity  
$v_*=16\sqrt{2}\Delta_0\Delta_I$ \cite{kolmogorov1study,fisher1937wave}.
In more detail,
$z(x-x_t,t)$ converges to the traveling wave solution $f(x)$, 
where the displacement of the wavefront is $x_t = v_*t + O(\ln t)$
\cite{bramson1983convergence,brunet2015exactly}.

In the forward tail of the travelling wave, ${f=1-\Delta f}$ with ${\Delta f\ll 1}$, and one may linearise the equation  in $\Delta f$ \cite{kolmogorov1study,fisher1937wave}. The linearised equation has solutions ${\Delta f \sim \exp({-\lambda (x-v(\lambda) t)})}$ for arbitrary $\lambda$,
with velocities ${v(\lambda) = 4 \Delta_0^2 \lambda + 32 \Delta_I^2 /\lambda}$.
When the initial ``domain wall'' is sufficiently well-localised, as in the above example, 
the velocity $v_*$ 
which emerges at late time
is the \textit{minimum}  of $v(\lambda)$  over all real $\lambda$, occurring for ${\lambda_* = 2\sqrt{2}\Delta_I/\Delta_0}$. 
For example, this can be seen by a graphical argument in which we plot $\ln \Delta f$ versus $x$ (see e.g. Ref.~\cite{munier2014lecture}).
If the initial condition has a right tail that relaxes to $z=1$ sufficiently slowly as $x\to+\infty$, then a faster travelling wave can result, but that case is not relevant to our discussion in this paper.
See the early parts of  Ref.~\cite{brunet2015exactly} for an overview of the asymptotic behaviour. 

Setting $y=x-vt$, the travelling wave solution obeys 
\begin{equation}
    -v f'(y) = 4\Delta_0^2f''(y)-16\Delta_I^2f(y)\left(1-f(y)^2\right)
\end{equation}
with $v=v_*$.
This has the form of the equation of motion for a particle moving in one dimension \cite{kolmogorov1study,fisher1937wave}
\begin{equation}
    m{\ddot {X}}=-V'(X)-\gamma\dot{X},
\end{equation}
with mass ${m=4\Delta_0^2}$, a  potential ${V(X)=4\Delta_I^2(1-X^2)^2}$, and a friction term with coefficient $\gamma=v$. Here  ${y= x-v_*t}$ has been interpreted as a fictitious time coordinate.
The boundary conditions are  such that the particle is  at the  local potential maximum $X=0$
in the distant past ($y\to-\infty$) and at the potential minimum 
$X=1$
in the distant future.
The velocity $v=v_*$ is the one for which the motion is critically damped. 
(For $v<v_*$ the motion would be underdamped, and the particle would oscillate around $X=1$, which is inconsistent with $z\leq 1$ in the original problem.)

\section{Linear approximation for bound state}

When the the velocity $v=0$, the fictitious dynamics settles on a stable domain-wall bound state configuration where $\theta(x)=\phi(x)$. When $v\neq 0$, it is no longer true that $\theta(x)=\phi(x)$, but for small velocities $\vert v\vert/v_c \ll 1$ we might expect the difference between $\theta$ and $\phi$ to be small. Here we show that this difference is of order $v$ at small $v$ (so that the separation between the two domain walls is also of order $v$ at small $v$).

Working with the moving-frame variables $\Theta$ and $\Phi$,
we can rewrite the equilibrium values  in the suggestive form
\begin{align}
\Theta(x)&=\theta_0(x)+\alpha(x) \\
    \Phi(x)&=\theta_0(x)-\alpha(x)
\end{align}
where $\alpha$ is small when $v$ is small (since the two fields $\Theta$ and $\Phi$ coincide when $v=0$).
Substituting into the steady state equations for $\Theta$ and $\Phi$ (see Eq.~\ref{eq:steadystate}) and expanding in both $v$ and $\alpha$, we find, at leading nontrivial order,
\begin{equation}
    -v\theta_0'=4\Delta_0^2\alpha''+\left(4\Delta_0^2(\theta_0')^2-8\Delta_I^2(1+\cos^2 2\theta_0)\right)\alpha,
\end{equation}
where $\theta_0$ is the ${v=0}$ solution. We see that   $\alpha$ is of order $v$, and we define
$\beta(x) \equiv\lim_{v\to 0}\alpha(x)/(v\Delta_0^{-1}\Delta_I^{-1})$ (recall that $v_c\propto\Delta_0\Delta_I$). For convenience we  also set ${\Delta_0=\Delta_I=1}$ by rescaling space and time as discussed in Sec.~\ref{sec:coherentstate}. Finally we substitute in the known analytic solution for $\theta_0$, to give
\begin{equation}
     \beta''(\tilde x)
    = (1+3~\mathrm{tanh}^2~2\tilde{x})\beta(\tilde x)
      -\frac{\mathrm{sech}~2\tilde{x}}{4}.
\end{equation}
This equation for $\beta(\tilde x)$
(which can be interpreted as the classical mechanics of a particle in a time-varying force)
can be solved numerically with the boundary conditions that ${\beta\to 0}$ as $\tilde x\to \pm\infty$. 

This solution can then be substituted into the action: this is an independent way to obtain the scaling function $g(v/v_c)$ up to quadratic order in its argument, 
\be
g(v/v_c)\simeq g(0)+k({v}/{v_c})^2,
\ee
with a numerical value for $k$ which we confirmed was roughly in agreement with the results in the main text. 

This approach only gives information for small $v$, but note that it only involves solving a single equation, rather than a separate equation for each $v$ as in the main text.

\section{Boundary conditions for the OTOC} \label{app:otoc-bcs}

The OTOC can be written as
\begin{equation}
    \frac{1}{\mathcal{N}}
    \sum_{n}
    \bra{n}
    \hat{U}^\dagger(T)\hat A ~\hat{U}(T) \hat{B}~
    \hat{U}^\dagger(T)\hat A ~\hat{U}(T) \hat{B}
    \ket{n}
\end{equation}
where the sum is over some complete basis of states $\ket{n}$.

For each state $\ket{n}$, the operator $\hat{B}$ is applied, then the state is time-evolved forward by $T$. Then $\hat{A}$ is applied and the state is time-evolved backwards in time by $T$. This process is repeated and then the final state is contracted with the original state $\ket{n}$. This is then summed for all possible $\ket{n}$.

This can calculated in the $N=2$ replica model as follows (assume replicas $a=1$
 and $a=3$ are ``forward'' replicas evolving under $\hat U$ and replicas $a=2$ and $a=4$ and ``backward'' replicas evoling under $\hat U^*$):
 
 Take the inital state of replica $1$ to be $\hat B\ket{n}$. This then evolves to $\hat{U}(T)\hat B\ket{n}$ at the final time. Then apply $\hat A$ at the final time to get $\hat A~\hat{U}(T)\hat B\ket{n}$. Then perform an ``identity'' index contraction on the final states of replicas $1$ and $2$ which ensures that the final state of replica $2$ is $(\hat A~\hat{U}(T)\hat B\ket{n})^*$ and therefore the initial state is $(\hat{U}^\dagger(T)\hat A~\hat{U}(T)\hat B\ket{n})^*$. Applying another index contraction to the initial states of replicas $2$ and $3$ to ensure that the initial state of replica $3$ is $\hat{U}^\dagger(T)\hat A~\hat{U}(T)\hat B\ket{n}$. Repeating this by applying $\hat B$ again at the initial time to replica $3$, applying $\hat A$ at the final time to replica $3$ and performing an index contraction to the final states of replicas $3$ and $4$ similarly ensures that the inital state of replica $4$ is $(\hat{U}^\dagger(T)\hat A~\hat{U}(T)\hat B~\hat{U}^\dagger(T)\hat A~\hat{U}(T)\hat B\ket{n})^*$.

 Finally, applying and index contraction to the inital states of replicas $1$ and $4$ gives the sum $\sum_n\bra{n}\hat{U}^\dagger(T)\hat A~\hat{U}(T)\hat B~\hat{U}^\dagger(T)\hat A~\hat{U}(T)\hat B\ket{n}$, which is simply the OTOC multiplied by the Hilbert space dimension $\mathcal{N}$.

 Therefore the OTOC can be calculated by choosing the following boundary conditions: at the initial time, we pair replica $1$ with $4$ and pair replica $2$ with $3$, as well as acting with $\hat B$ on replicas $1$ and $3$. At the final time, we pair replica $1$ with $2$ and pair replica $3$ with $4$, as well as acting with $\hat A$ on replicas $1$ and $3$.

 The pairing at the initial time is the $N=2$ ``swap'' contraction, and corresponds to the state $\mid\rightarrow\ra$ in the spin model. However, applying $\hat B$ to replicas $1$ and $3$ swaps the spin direction $\mid\leftarrow\ra$ on every site that $\hat B$ contains a Majorana operator.

Similarly, the pairing at the final time is the $N=2$ ``identity'' contraction, and corresponds to the state $\mid\uparrow\ra$ in the spin model. Applying $\hat A$ to replicas $1$ and $3$ swaps the spin direction $\mid\downarrow\ra$ on every site that $\hat A$ contains a Majorana operator.
 
\section{Markov process for evolving operator} \label{app:markov}

In order to derive the master equation for the time evolution of $\overline{a_S(t)^2}$ it will be convenient to Trotterize the time-evolution operator into random unitaries  of the form $\exp(\eta\hat\gamma_i\hat\gamma_{i+1})$ and $\exp(i\eta'\hat\gamma_i\hat\gamma_{i+1}\hat\gamma_{i+2}\hat\gamma_{i+3})$, where $\eta$ and $\eta'$ are now Gaussian random variables with mean zero and finite variances $\Delta_0^2\delta t$ and $\Delta_I^2\delta t$ respectively. In order to reproduce the continuous time model in the limit $\delta t\to 0$, in a fixed time interval $\Delta t$ each type of unitary gate should be applied $\Delta t/\delta t$ times (for each $i$).

We can now ask how Hermitian operators in the spin chain evolve under the action of these unitary gates. To do this, we only need to know how arbitrary products of Majorana operators evolve, given that any Hermitian operator in the Majorana chain is a linear of combination of such products (\ref{eq:opexpansion}). Here we restrict to products of an even number of Majoranas.

For each type of gate we get a discrete Heisenberg equation of motion for the operator $\hat A$
\begin{align} \notag
    e^{-\eta\hat\gamma_i\hat\gamma_{i+1}}
    \hat A
    e^{\eta\hat\gamma_i\hat\gamma_{i+1}}
    \approx
    \hat A + \eta[\hat A,\hat\gamma_i\hat\gamma_{i+1}] \\
    -\eta^2\hat\gamma_i\hat\gamma_{i+1}\hat A\hat\gamma_i\hat\gamma_{i+1}
    -\eta^2\hat A
    +O(\eta^3)
\end{align}
\begin{align} \notag
    e^{-i\eta'\hat\gamma_i\hat\gamma_{i+1}\hat\gamma_{i+2}\hat\gamma_{i+3}}
    \hat A &
    e^{i\eta'\hat\gamma_i\hat\gamma_{i+1}\hat\gamma_{i+2}\hat\gamma_{i+3}}
    \approx
    \hat A \\&+ i\eta'[\hat A,\hat\gamma_i\hat\gamma_{i+1}\hat\gamma_{i+2}\hat\gamma_{i+3}] \notag \\
    & +\eta'^2\hat\gamma_i\hat\gamma_{i+1}\hat\gamma_{i+2}\hat\gamma_{i+3}\hat A
    \hat\gamma_i\hat\gamma_{i+1}\hat\gamma_{i+2}\hat\gamma_{i+3} \notag\\
     &-\eta'^2\hat A +O(\eta'^3)
\end{align}

Let's consider the commutators in turn:

If $\hat A$ is a product of Majorana operators, then $\hat\gamma_i\hat\gamma_{i+1}$ commutes with $\hat A$ unless $\hat A$ contains exactly one of $\hat\gamma_i$ or $\hat\gamma_{i+1}$. If it does contain exactly one of these, then $[\hat A,\hat\gamma_i\hat\gamma_{i+1}]$ is also proportional to a product of Majorana operators similar to $\hat A$, but $\hat\gamma_i$ replaced by $\hat\gamma_{i+1}$ or vice versa. This term therefore causes Majorana operators to ``hop'' to their nearest neighbours. The proportionality factor is  $+2$ if $\hat\gamma_i$ is replaced by $\hat\gamma_{i+1}$, and is $-2$ if $\hat\gamma_{i+1}$ is replaced by $\hat\gamma_i$ (the signs won't be important in the end because the commutator is multiplied by $\eta$ which has a random sign). So, in the case where $\hat A$ does not commute with $\gamma_i\gamma_{i+1}$,
\ba\label{eq:Aevo1}
e^{-\eta\hat\gamma_i\hat\gamma_{i+1}}
    \hat  A
    e^{\eta\hat\gamma_i\hat\gamma_{i+1}}
    \approx
 (1-2 \eta^2)   A \pm 2 \eta A|_{\gamma_i\leftrightarrow \gamma_{i+1}}.
\end{align}

Similarly, $\hat A$ commutes with $\hat\gamma_i\hat\gamma_{i+1}\hat\gamma_{i+2}\hat\gamma_{i+3}$ unless $\hat A$ contains exactly one or three of the Majorana operators from $i$ to $i+3$. If it doesn't commute, then $[\hat A,\hat\gamma_i\hat\gamma_{i+1}\hat\gamma_{i+2}\hat\gamma_{i+3}]$ is a product of Majoranas similar to $\hat A$, but the Majorana operators from $i$ to $i+3$ which did appear $\hat A$ no longer appear, and the ones which did not appear in $\hat A$ now do appear. This means that either one Majorana operator is replaced by three, or three Majorana operators are replaced by one. Unlike the quadratic terms which simply caused the Majorana operators to hop while conserving their number, the quartic terms allow Majorana operators to be created and destroyed while conserving parity. There is also a factor of $\pm 2$, with the sign depending on which Majorana operators are replaced. This gives an evolution analogous to (\ref{eq:Aevo1}).

We can summarise this using the following matrices (recall that $S$, $S'$ refer to strings of Majorana operators, see Eq.~\ref{eq:Gammadef})
\begin{align}
    M_{SS'}^{(i)}&=\frac{1}{\mathcal{N}}
    \Tr\left[\hat\Gamma_S[\hat \Gamma_{S'},\hat\gamma_i\hat\gamma_{i+1}]\right] \\
    {{M'}_{SS'}^{(i)}}&=\frac{1}{\mathcal{N}}
    \Tr\left[\hat\Gamma_S[\hat \Gamma_{S'},\hat\gamma_i\hat\gamma_{i+1}\hat\gamma_{i+2}\hat\gamma_{i+3}]\right]
\end{align}
so $M^{(i)}_{SS'}={0, \pm 2}$ gives the coefficient of $\hat\Gamma_S$ coming from the commutator $[\hat \Gamma_{S'},\hat\gamma_i\hat\gamma_{i+1}]$ and ${{M'}_{SS'}^{(i)}}$ does the same for $[\hat\Gamma_{S'},\hat\gamma_i\hat\gamma_{i+1}\hat\gamma_{i+2}\hat\gamma_{i+3}]$.
Note that, if $S$ can be reached from $S'$ by application of a gate, there is only one possible gate which effects this transition. So if we define
\be
M_{SS'} = \sum_i M^{(i)}_{SS'},
\ee
and similarly for $M'$, then we again have 
$M_{SS'}=0,\pm 2$ and $M'_{SS'}=0,\pm 2$.

Now we assume that, for a particular realisation of the noise $\eta$ and $\eta'$, we know the coefficients $a_S(t)$ in Eq.~\ref{eq:opexpansion} at some time $t$. 
We may define the Trotterization such that, in each  $\delta t$ interval, we apply each kind of unitary gate once for each $i$. 
Then we can write the new coefficients after one more timestep as
\begin{align} \notag
    &a_S(t+\delta t)\approx \\
    &\left(1-\frac{1}{2}\sum_{S'}M_{S'S}^2\eta_{S'S}^2-\frac{1}{2}\sum_{S'}{M'}_{S'S}^2{\eta'}_{S'S}^2\right)a_S(t) \notag\\
    &+\sum_{S'}M_{SS'}\eta_{SS'}a_{S'}(t)
    +\sum_{S'}M'_{SS'}\eta'_{SS'}a_{S'}(t)\notag
\end{align}
where $\eta_{SS'}$ and $\eta'_{SS'}$ denote the noise terms which multiply the commutators which produce $\hat\Gamma_S$ from $\hat\Gamma_{S'}$, and in the first line we have used the fact that $M_{S'S}$ and ${M'}_{S'S}$ are $\pm 2$ whenever they are non-zero. There is only at most one way to produce a given $\hat\Gamma_S$ from a given $\hat\Gamma_{S'}$, so the noise variables $\eta_{SS'}$ and $\eta'_{SS'}$ 
for the nonvanishing terms are all independent and are simply a relabelling of
the $\eta_i$ and the $\eta'_i$.

We can then write the noise-averaged square of ${a_S(t+\delta t)}$ conditioned on the previous coefficents $a_S(t)$, which just involves  averaging over $\eta_{SS'}$ and $\eta'_{SS'}$ which are independent from $a_S(t)$:
\begin{align} \notag
    &\mathbb{E}(a_S(t+\delta t)^2|a_S(t))\approx~ \\
    &\left(1-\Delta_0^2\delta t\sum_{S'}M_{S'S}^2-\Delta_I^2\delta t\sum_{S'}{M'}_{S'S}^2\right)a_S(t)^2 \notag\\
    &+ \Delta_0^2\delta t\sum_{S'}M^2_{SS'}a_{S'}(t)^2
    + \Delta_I^2\delta t\sum_{S'}M'^2_{SS'}a_{S'}(t)^2+\cdots,
\end{align}
where the only second order terms that survive the averaging are ones where the same $\eta$ or $\eta'$ appears twice.

To get the unconditional expectation value, we integrate with respect to the coefficients $a_S(t)$ weighted by their probability $P(\{a_S(t)\})$ to get
\begin{align} \notag
    &\overline{a_S(t+\delta t)^2}\approx \\
    &\left(1-\Delta_0^2\delta t\sum_{S'}M_{S'S}^2-\Delta_I^2\delta t\sum_{S'}{M'}_{S'S}^2\right)\overline{a_S(t)^2} \notag\\
    &+ \Delta_0^2\delta t\sum_{S'}M^2_{SS'}\overline{a_{S'}(t)^2}
    + \Delta_I^2\delta t\sum_{S'}M'^2_{SS'}\overline{a_{S'}(t)^2} + \cdots.
\end{align}
or more concisely
\begin{equation}
    \overline{a_S(t+\delta t)^2}\approx~
    \overline{a_S(t)^2}
    + \delta t\sum_{S'}R_{SS'}\overline{a_{S'}(t)^2}
\end{equation}
where we have defined
\begin{align}    R_{SS'} = & \Delta_0^2M^2_{SS'}+\Delta_I^2M'^2_{SS'}
    \\  & -\delta_{SS'}\sum_{S''}\left(\Delta_0^2 M_{S''S'}^2 +\Delta_I^2 M_{S''S'}'^2\right).
\end{align}
We can view  $R$ as the transition matrix  for a stochastic particle hopping problem, with occupied sites corresponding to the sites represented in the string $S$.
 $R_{SS'}$ is equal to $4\Delta_0^2$ whenever $S$ and $S'$ differ only by a single particle hopping to an unoccupied nearest neighbour, and is equal to $4\Delta_I^2$ when $S$ and $S'$ differ only by reversing the occupancies of a set of four contiguous sites that contains an  odd number of particles.

%\section{Numerical convergence of fictitious dynamics} \label{app:convergence}

\bibliographystyle{apsrev4-1}
\bibliography{noisymajorana2.bib}

\end{document}